\documentclass[%
reprint,
superscriptaddress,
nofootinbib,
amsmath,amssymb,
aps,
floatfix,
]{revtex4-2}

\usepackage{dcolumn}
\usepackage{bm}
\usepackage[english]{babel}
\usepackage{graphicx}
\usepackage{xspace}
\usepackage{multirow}
\usepackage{booktabs}
\usepackage[caption=false]{subfig}
\usepackage{enumerate}
\usepackage[compat=1.1.0]{tikz-feynman}
\usepackage{tikz}
\usepackage[colorlinks=true, allcolors=magenta]{hyperref}
\usepackage{floatrow}

\newif\ifcomments
\commentstrue 

\newcommand{\Beta}    {\ensuremath{\beta}\xspace}
\newcommand{\nnbb}    {\ensuremath{2\nu\beta\beta}\xspace}
\newcommand{\onbb}    {\ensuremath{0\nu\beta\beta}\xspace}
\newcommand{\Ca}      {\ensuremath{^{48}}Ca\xspace}
\newcommand{\Ge}      {\ensuremath{^{76}}Ge\xspace}
\newcommand{\Zr}      {\ensuremath{^{96}}Zr\xspace}
\newcommand{\Xe}      {\ensuremath{^{136}}Xe\xspace}
\newcommand{\Se}      {\ensuremath{^{82}}Se\xspace}
\newcommand{\Te}      {\ensuremath{^{130}}Te\xspace}
\newcommand{\Mo}      {\ensuremath{^{100}}Mo\xspace}
\newcommand{\Cd}      {\ensuremath{^{116}}Cd\xspace}
\newcommand{\Nd}      {\ensuremath{^{150}}Nd\xspace}

\newcommand{\bbHL}    {\ensuremath{T_{1/2}^{2\nu}}\xspace}

\newcommand{\qbb}     {\ensuremath{Q_{\beta\beta}}\xspace}
\newcommand{\gerda}   {\textsc{Gerda}\xspace}
\newcommand{\legend}  {LEGEND\xspace}
\newcommand{\majorana}{\textsc{Majorana Demonstrator}\xspace}

\newcommand{\alv}     {\ensuremath{\smash{\mathring{a}_\text{of}^{(3)}}}\xspace}
\newcommand{\sinT}    {\ensuremath{\sin^2\theta}\xspace}

\newcommand{\gfermi}  {G\ensuremath{_F}\xspace}
\newcommand{\nusi}    {\ensuremath{\nu}SI\xspace}
\newcommand{\nnsibb}  {\ensuremath{2\nu}SI\ensuremath{\beta\beta}\xspace}

\newcommand{\Jbb}    {J\ensuremath{\beta\beta}\xspace}
\newcommand{\JJbb}   {JJ\ensuremath{\beta\beta}\xspace}
\newcommand{\phibb}  {\ensuremath{\phi\beta\beta}\xspace}

\newcommand{\Ztwo}    {\ensuremath{Z_2}\xspace}

\newcommand{\cl}   {\ensuremath{\text{C.L.}}\xspace}
\newcommand{\ci}   {\ensuremath{\text{C.I.}}\xspace}

\begin{document}


\title{Probing Beyond the Standard Model Physics with Double-beta Decays}

\author{E.~Bossio}
\affiliation{Physik Department, Technische  Universit{\"a}t M{\"u}nchen, Germany}
\email{elisabetta.bossio@tum.de}

\author{M.~Agostini}
\affiliation{Department of Physics and Astronomy, University College London, London, UK}
\email{matteo.agostini@ucl.ac.uk}

\date{\today}

\begin{abstract}
Nuclear double-beta decays are a unique probe to search for new physics beyond the Standard Model. Still-unknown particles, non-standard interactions, or the violation of fundamental symmetries would affect the decay kinematic, creating detectable and characteristic experimental signatures. In particular, the energy distribution of the electrons emitted in the decay gives an insight into the decay mechanism and has been studied in several isotopes and experiments. No deviations from the prediction of the Standard Model have been reported yet. However, several new experiments are underway or in preparation and will soon increase the sensitivity of these beyond the Standard Model physics searches, exploring uncharted parts of the parameter space. This review brings together phenomenological and experimental aspects related to new-physics searches in double-beta decay experiments, focusing on the testable models, the most-sensitive detection techniques, and the discovery opportunities of this field. 
\end{abstract}

\maketitle
\begingroup
  \hypersetup{hidelinks}
  \tableofcontents
\endgroup

\section{Introduction}\label{sec:intro}

Double-\Beta decays are nuclear transitions in which an isotope changes its atomic number $Z$ by two units, while maintaining constant its mass number $A$: 
\begin{equation}
 (A, Z) \rightarrow (A,Z+2) +  ....
\end{equation}
This process can occur only if the conversion of two protons into two neutrons leads to a more bounded nuclear configuration, with a positive Q-value defined as $\qbb \approx M(A,Z) c^2 - M(A,Z+2) c^2$~\cite{Detwiler:2022tfk}, where $M$ refers to the mass of the isotopes. Different theory models predict different decay final states. However, in general, they all envision two electrons to conserve the electric charge, along with neutral particles.

\begin{figure}[tbp]
\begin{tikzpicture}
    \draw[thick,-Straight Barb] (0,0) -- (6.,0) ;
    \draw[thick,-Straight Barb] (0,0) -- (0,4.5) node[rotate=90,anchor=south east] {M(A,Z)};
    \draw (3 cm,1pt) -- (3 cm,-1pt) node [anchor=north] {Z};
    \draw (4 cm,1pt) -- (4 cm,-1pt) node[anchor=north] {Z+1};
    \draw (5 cm,1pt) -- (5 cm,-1pt) node[anchor=north] {Z+2};
    \draw (2 cm, 1pt) -- (2 cm, -1pt) node [anchor=north] {Z-1};
    \draw (1 cm, 1pt) -- (1 cm, -1pt) node [anchor=north] {Z-2};
    \draw (0,4) parabola bend (3,0) (6,4) ;
    \draw (5.5,2.25) node [anchor=north west] {even/even};
    \draw[gray] (1,4) node [anchor=south west] {odd/odd}  parabola bend (3,2) (5,4) ;
    \fill[black] (3,0) circle (2pt);
    \fill[black] (1, 1.77778) circle (2pt);
    \fill[black] (5, 1.77778) circle (2pt);
    \fill[gray] (2, 2.5) circle (2pt);
    \fill[gray] (4, 2.5) circle (2pt);
    \draw[thick,-Straight Barb] (1, 1.777778) -- (2.9,0.1) node [pos=0.4, right, inner sep= 8pt] {$\beta^-\beta^-$};
    \draw[thick,-Straight Barb] (1, 1.777778) -- (1.9,2.4) node [pos=0.4, right, inner sep= 8pt] {$\beta^-$};
    \end{tikzpicture}
\caption{Mass parabolas of nuclear isobars with even A. Due to the pairing term in the semi-empirical mass formula, $\beta^-$ transitions of even/even nuclei to their odd/odd isobaric neighbor can be energetically forbidden, whereas, in a second-order process, $\beta^-\beta^-$ decay is allowed. \label{fig:mass_parabola}}
\end{figure}
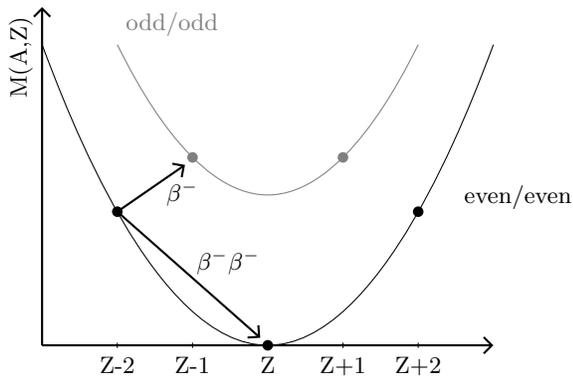

These decays are a second-order weak process. Thus, they can be observed only in isotopes for which the otherwise dominant first-order transitions --- and in particular, the single-$\beta$ decay --- are strongly suppressed. 
This is the case for a limited number of even-even nuclei for which the single-\Beta decay is energetically forbidden, as the attractive nuclear pairing interaction makes them more bound than their odd-odd neighbors but less than their even-even second-neighbors  (see \figurename~\ref{fig:mass_parabola}). Alternatively, candidate isotopes for the observation of double-\Beta transitions are also those for which the single-\Beta decay is suppressed by the large mismatch between the total angular momentum of the initial and final nuclei~\cite{Alanssari:2016jtf}.

Double-\Beta decays were postulated in  1935~\cite{Goeppert1935} by Maria Goeppert-Maier, who pointed out how these transitions could proceed through the ``simultaneous emission of two electrons and two (anti)neutrinos'':
\begin{equation}
    (A, Z) \rightarrow (A,Z+2) + 2e^- + 2\overline{\nu}
    \label{eq:nnbb-eq}
\end{equation}
In this case, the decay would ``appear as the simultaneous occurrence of two transitions, each of which does not fulfill the law of conservation of energy separately''. This final state is the only one allowed by the Standard Model of particle physics and is typically referred to as ``two-neutrino double-\Beta'' (\nnbb) decay. 

The discovery of double-\Beta transitions traces back to the observation of the decay daughter isotope in a geochemical experiment with \Te~\cite{Inghram1950}. 
Only 40 years later, double-\Beta decays were observed in real-time through a calorimetric measurement of the energy released by the electron emitted in the decay of \Se~\cite{Elliott1987b}. The electron summed energy distribution was found to have a continuous shape well compatible with the expectations for Goeppert-Maier's \nnbb decay (see \figurename~\ref{fig:experimental_signature_2nbb}). 
To date, double-\Beta transitions compatible with the \nnbb final state have been observed in 9 isotopes with half-life values in the range of $10^{18}-10^{24}$\,yr~\cite{Barabash2020}, making it one of the rarest processes ever measured by humankind. 

Theories beyond the Standard Model (BSM) predict a variety of new, additional final states involving the emission of electrically neutral exotic particles, which would affect the decay kinematics and alter the energy and angular distributions of the electrons emitted in the decay. Similarly, also the violation of fundamental symmetries would affect the decay kinematic. 
This is why double-\Beta decays have been a unique window to new physics for several years. However, despite many past and recent experiments, no deviations from the standard Model's \nnbb-decay expectations have yet been observed. 

Double-\Beta decay searches have been historically carried out in experiments primarily built to study a hypothesized final state with only two electrons (i.e. without neutrinos), in which the summed electron energy is expected to be precisely equal to \qbb rather than a continuous distribution (see \figurename~\ref{fig:experimental_signature_2nbb}). Such ``neutrinoless double-\Beta'' (\onbb) decay is indeed predicted by our leading theories explaining the matter-antimatter asymmetry and origin of neutrino masses. Its search has been an indisputable priority of the particle-physics community for the last two decades. Currently, ton-scale double-\Beta-decay experiments are under preparation with the ultimate goal of increasing the sensitivity to \onbb-decay half-life values by two orders of magnitudes compared to current constraints, up to $10^{28}$\,yr. We refer the reader to Ref.~\cite{Agostini:2022zub} for a recent review on \onbb decay.

Thanks to their planned ultra-low background level and huge target mass,
the next-generation double-\Beta decay experiments will provide high-precision measurements of the \nnbb-decay, enabling new opportunities to discover new final states. 
This motivates the timing of our work.
Indeed, although several review articles have discussed double-\Beta decay searches in recent years, they have typically focused on the \nnbb-decay final state. For instance, we would like to highlight Ref.~\cite{Barabash:2011mf} that gives an excellent review of \nnbb-decay's history and \cite{Saakyan:2013yna} for a recent summary of the field. Differently from past works, our review is the first one focused on the BSM searches, covering their theoretical and experimental aspects, current constraints, and ongoing endeavor to improve the experimental sensitivities. Future \onbb decay experiments with leading sensitivity will measure only the summed energy of the two final state electrons -- aiming to distinguish the \onbb decay peak at \qbb from the continuous \nnbb decay distribution -- without having access to the single electron energy or electron angular correlation. Therefore, this review focuses on how BSM physics would affect the summed electron energy distribution and how this can be exploited to search for BSM double-\Beta decays.  

The manuscript is organized as follows. We first briefly discuss the Standard Model \nnbb decay in Sec.~\ref{sec:nnbb} for completeness. We then review in
Sec.~\ref{sec:models_and_motivations} the phenomenology of double-\Beta transitions mediated by Beyond-Standard-Model processes. Section~\ref{sec:analysis_methods} is dedicated to their experimental signatures exploited by experiments. It also contains an original contribution to the field, which is the first derivation of analytical formulas describing the experimental sensitivity as a function of key experimental parameters and systematic uncertainties. 
Finally, Sec.~\ref{sec:experimental_results} summarises the latest experimental results and the prospects for next-generation experiments.

\section{Standard-Model allowed double-\Beta decay}
\label{sec:nnbb}

Double-\Beta transitions can be uniquely identified by the production of the daughter nucleus. With this principle, the double-\Beta decay of \Te was first detected in 1950 with the first geochemical experiment. The detection of an excess of $^{130}$Xe was proof of the double-\Beta decay of the initial nucleus and allowed a first determination of its half-life~\cite{Inghram1950}. Even if this result was initially not considered seriously, it represents the first observation of \nnbb decay, as it became clear after 15--20\,yr. 

In fact, after both the \nnbb and \onbb decays were proposed (1935--1939), the first estimate of the half-life of the two processes strongly favored the \onbb decay. This was predicted with a half-life of the order of $10^{15}$\,yr, whereas a half-life of about $10^{21}$\,yr was predicted for the Standard-Model allowed \nnbb decay. Only the discovery of parity violation in 1957 and the determination of the V--A nature of weak interaction made clear that the probability of \onbb decay had to be much smaller than that of \nnbb decay.  

In the following years, new geochemical experiments were performed, confirming the observation of the \nnbb decay of \Te~\cite{TakaokaOgata1966, Kirsten1967} and observing for the first time the \nnbb decay of \Se~\cite{Kirsten1967} and $^{128}$Te~\cite{Manuel1975}.     

\begin{figure}[tbp]
    \centering
    \includegraphics[width=\textwidth]{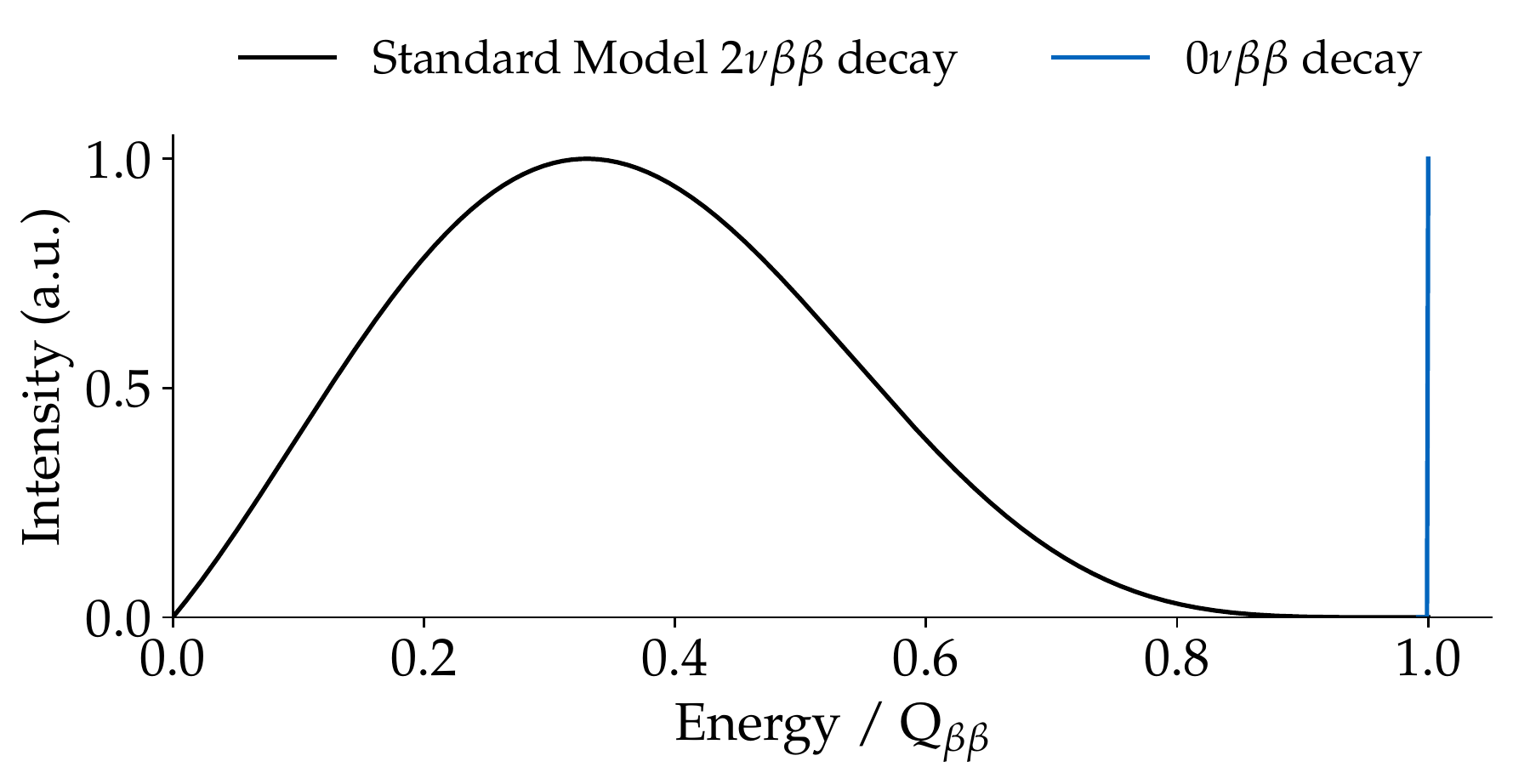}
    \caption{Summed electron energy distribution for the SM \nnbb decay and the lepton number non-conserving \onbb decay. An infinite energy resolution is assumed, and an arbitrary normalization is used for illustrative purposes.}
    \label{fig:experimental_signature_2nbb}
\end{figure}

A key milestone in the history of double-\Beta decays is the theoretical work of Doi, Kotani, and Takasugi in 1985~\cite{Doi1985}, who first computed the energy and angular distributions of the electrons emitted in double-\Beta decays, providing a clear signature to distinguish the Standard-Model allowed \nnbb decay from BSM double-\Beta decays. In the same work, \figurename~\ref{fig:experimental_signature_2nbb} appeared for the first time. 

Following these theory developments, the first direct observation of a \nnbb decay was made in 1987 using a \Se time projection chamber, which could measure the summed energy distribution of the two electrons  and extract the \Se \nnbb-decay half-life based on 36 observed events. This was a turning point in the history of \nnbb decays, and in the following ten years \nnbb decay was directly observed in seven isotopes: \Ge~\cite{vasenko1990}, \Mo~\cite{EJIRI199117, Elliott_1991, PhysRevC.56.2451, PhysRevD.51.2090}, \Se~\cite{Elliot1992, ARNOLD1998209}, \Cd~\cite{Ejiri1995, NEMO:1995asg, Danevich:1995np, nemo1996}, \Ca~\cite{PhysRevLett.77.5186}, \Nd~\cite{PhysRevC.56.2451}, \Zr~\cite{ARNOLD1999299}. In addition to the summed electrons' energy distribution, which was measured for all the above-mentioned isotopes, the NEMO-2 experiment also measured the single electron energy and the angular distributions of the electrons for \Mo, \Cd, \Se, and \Zr~\cite{PhysRevD.51.2090, NEMO:1995asg, nemo1996, ARNOLD1998209, ARNOLD1999299}. In the same years, the \nnbb decay of $^{238}$U was first observed in a radiochemical experiment~\cite{PhysRevLett.67.3211}. 

By the end of the 20th century, the energy and half-life of \nnbb decay had been measured for 10 isotopes, considering direct, geochemical, and radiochemical experiments. In addition, in the last two decades, direct observation of the \nnbb decay of \Te and \Xe were reported in 2010~\cite{Barabash2011} and 2011~\cite{EXO-200:2011xzf}, respectively.
Table~\ref{tab:nnbb-isotopes} summarises the double-\Beta decaying isotopes for which the \nnbb decay was directly observed. 

The precision with which experiments were able to determine the half-life of \nnbb decays has increased over the years, with modern experiments reaching a percent-level precision for most of the isotopes. The most precise measurement of the \nnbb decay half-life are summarised in Table~\ref{tab:nnbb-isotopes}.      

\begin{table*}[tbp]
    \centering
    \renewcommand{\arraystretch}{1.2}
    \begin{tabular}{cccccc}
    \toprule
         Isotope  & First & Ref. & Half-life & Experiment & Ref.   \\
           & observation & & & \\ \midrule
         \Ca $\rightarrow \, ^{48}$Ti & 1996 & \cite{PhysRevLett.77.5186} & $( 6.4 ^{+1.4}_{-1.1}) \cdot 10^{19}$ & NEMO-3 & \cite{Arnold2016} \\
         \Ge $\rightarrow \, ^{76}$Se & 1990 & \cite{vasenko1990} & $(2.022 \pm 0.041) \cdot 10^{21}$ & GERDA & \cite{gerda2022c} \\
         \Se $\rightarrow \, ^{82}$Kr & 1987 & \cite{Elliott1987b} & $(8.60 ^{+0.19}_{-0.13}) \cdot 10^{19}$ & CUPID-0 &  \cite{Azzolini2019b} \\
         \Zr $\rightarrow \, ^{96}$Mo & 1999 & \cite{ARNOLD1999299} & $(2.35 \pm 0.21) \cdot 10^{19}$ & NEMO-3 & \cite{Argyriades2009}\\
         \Mo $\rightarrow \, ^{100}$Ru & 1991 & \cite{EJIRI199117, Elliott_1991} & $(7.12 ^{+0.21}_{-0.17}) \cdot 10^{18}$ & CUPID-Mo & \cite{Armengaud2020} \\
         \Cd $\rightarrow \, ^{116}$Sn & 1995 & \cite{Ejiri1995, NEMO:1995asg, Danevich:1995np} & $(2.63 ^{+0.11}_{-0.12}) \cdot 10^{19}$  & Aurora & \cite{Barabash2018} \\
         \Te $\rightarrow \, ^{130}$Xe & 2010 & \cite{Barabash2011} & $(7.71 ^{+0.14}_{-0.16}) \cdot 10^{20}$ & CUORE & \cite{Adams2021} \\
         \Xe $\rightarrow \, ^{136}$Ba & 2011 & \cite{EXO-200:2011xzf} & $(2.165 \pm 0.063) \cdot 10^{21}$ & EXO-200 & \cite{Albert2014}\\
         \Nd $\rightarrow \, ^{150}$Sm & 1997 & \cite{PhysRevC.56.2451}& $(9.34 ^{+0.66}_{-0.64}) \cdot 10^{18}$ & NEMO-3 & \cite{Arnold2016b}\\
         \bottomrule
    \end{tabular}
    \caption{Direct observations of \nnbb decays. The year of the first observation is indicated for each isotope, together with the most precise half-life determination. The shown uncertainty is the sum in quadrature of the statistical and systematic uncertainties, when available.}
    \label{tab:nnbb-isotopes}
\end{table*}

In addition to the Goeppert-Maier's \nnbb decay (Equation~\ref{eq:nnbb-eq}) or $\beta^- \beta^-$ transition, depending on the relative numbers of protons and neutrons in the nucleus, three additional second-order transitions are allowed in the Standard Model:
\begin{subequations}
\begin{align}
    (\beta^+ \beta^+)&: \; (A, Z) \rightarrow (A,Z-2) + 2e^+ + 2\nu  \\
    (ECEC)&: \; (A,Z) + 2e^- \rightarrow (A,Z-2) + 2\nu  \\ 
    (EC\beta^+)&: \; (A,Z) + e^- \rightarrow (A,Z-2) + e^+ + 2\nu \, .
\end{align}
\end{subequations} 
The energy released in the three processes listed above is smaller compared to the $\beta^- \beta^-$ decay in Equation~\ref{eq:nnbb-eq}. Consequently, these processes have lower probabilities compared to the $\beta^- \beta^-$ decay due to the smaller phase space, and experimentally they are much more challenging to observe. In the following, we will always refer to the $\beta^- \beta^-$ process as double-\Beta decay. 
  
In nature, there are 35 isotopes that can undergo \nnbb decay, and 34 more that can undergo $\beta^+ \beta^+$, $ECEC$, and $EC\beta^+$~\cite{TRETYAK200283}. In fact, not all of them fulfill the experimental requirements, {\it e.g.}, high isotopic abundance, large \qbb value, and compatibility with experimental technologies. To date, only the nine isotopes listed in Table~\ref{tab:nnbb-isotopes} were used in direct search experiments. 

The rate of \nnbb decay can be calculated following Fermi's golden rule for \Beta decay. To a good approximation, the kinematic part (phase space of the leptons emitted in the decay) and the nuclear part (matrix element responsible for the transition probability between the two nuclear states) can be factorized as:
\begin{equation}
    \Gamma^{2\nu} = [\bbHL]^{-1} = |\mathcal{M}^{2\nu}|^2 \, \mathcal{G}^{2\nu}(\qbb,Z)   \;,
\end{equation}
where $\mathcal{G}^{2\nu}$ is the phase-space factor and is obtained by integrating over the phase space of the four leptons, and $\mathcal{M}^{2\nu}$ is the nuclear matrix element (NME) and deals with the nuclear structure of the transition. While the phase-space factor can be calculated exactly, the NME is much more difficult to evaluate and relies on nuclear structure models. 

Experiments measure the distribution of the summed kinetic energy of the two electrons ($K$):
\begin{equation}\label{eq:nnbb_decay_rate_intro}
    \frac{d\Gamma^{2\nu}}{dK} = |\mathcal{M}^{2\nu}|^2 \, \frac{d\mathcal{G}^{2\nu}}{dK} \;.
\end{equation}
To a first approximation, the shape of this distribution is determined only by the phase space. The contribution of the NME to it is small and primarily affects the absolute value of the transition probability. 

The summed energy of the two electrons emitted in the \nnbb decay is continuously distributed between 0 and the endpoint at the \qbb value due to the neutrinos escaping the detector and carrying away part of the energy\footnote{To be precise, the maximum energy is \qbb minus the mass of the two emitted neutrinos. This is typically neglected as Q-values are at the MeV-energy scale, while neutrino masses are smaller than the eV-energy scale~\cite{KATRIN:2021uub, DES:2021wwk}.}. This is shown in Figure~\ref{fig:experimental_signature_2nbb}, compared to the \onbb decay, for which a $\delta$ function at \qbb is expected because all the transition energy goes into the kinetic energy of the two electrons. 

In conclusion, we have seen that even if the first observation of \nnbb decay is commonly traced back to 1950 and the first geochemical experiment with \Te, it took many years for the community to convince themselves that the production of the daughter nucleus observed in this first experiment was exactly the result of Goeppert-Maier's \nnbb decay. To date, precision measurements of the \nnbb decay of several isotopes and the agreement between the distribution of multiple observables (electrons summed energy, single electron energy, and angular distributions) with their theoretical prediction are striking evidence for the \nnbb decay and exclude a large part of the parameter space for many BSM theories. 
 
\section{Beyond-Standard-Model physics in double-\Beta decay}\label{sec:models_and_motivations}

This section reviews double-\Beta decays in the presence of BSM  physics. 

Suppose new particles were involved in a double-\Beta decay, or any new physics affected the phase space of the two electrons emitted in the \nnbb decay. In that case, the expected distribution of the summed electron energy would differ from that predicted for the SM \nnbb decay (equation~\ref{eq:nnbb_decay_rate_intro}, figure~\ref{fig:experimental_signature_2nbb}). This is the main feature used to search for these BSM decays in the experimental data. Single electron energy distributions and electron angular distributions are also primarily determined by the phase space, and they are characteristic of the physics model and decay final state. Experiments that can also measure these distributions can strongly enhance their sensitivity in distinguishing the SM allowed \nnbb decay from BSM double-\Beta decays. 
This review mainly discusses the first aspect, given that the calorimetric measurement approach pursued in the next generation of \onbb decay experiments with leading sensitivities gives access only to the summed electron energy distribution. 

We classify the models into three groups. The first one contains those predicting the existence of new particles --- either bosons or fermions --- which are emitted in the decay, replacing one or both of the \nnbb's neutrinos. The second one includes theories in which fundamental symmetries such as Lorentz covariance or Pauli's exclusion principle are violated. The last group covers non-standard interactions, like right-handed leptonic currents and strong neutrino self-interactions.

\subsection{New particles}\label{subsec:new_particles}

\subsubsection{Bosons}\label{subsubsec:majorons}

In the early 80s, an attractive approach to the neutrino mass problem was considered in which the neutrinos are Majorana particles with small masses arising from the spontaneous breakdown of the global B-L symmetry. In these models, a massless Goldstone boson should exist, which was called the "Majoron". Several realizations of this idea were proposed, mainly differing by the weak isospin ($I$) properties of the Majoron and leading to different phenomenology. 

The first model was proposed  by Chikashige, Mohapatra, and Peccei~\cite{Chikashige1981, Chikashige1980}. In this model, the Majoron arises from a Higgs singlet ($I=0$) and gives rise to small neutrino masses via the "see-saw" mechanism. On the other hand, the Majoron coupling to neutrino is so small that it would be very hard to test it through laboratory experiments. Shortly after, Gelmini and Roncadelli proposed a model in which the Majoron arises from a Higgs triplet ($I=1$)~\cite{Gelmini1981}. In this second case, a stronger coupling to neutrinos would be possible. A third case was considered later in 1987, in relation to solar neutrino oscillations, where the Majoron arises from a Higgs doublet ($I=1/2$), possibly leading to strong coupling to neutrinos~\cite{SANTAMARIA1987423, PhysRevLett.60.397, PhysRevD.39.1780}. 

If the Majoron Yukawa coupling to neutrinos were sufficiently strong ($\sim 10^{-5}-10^{-3}$), this would have interesting consequences for particle physics, astrophysics, and cosmology. Already in 1981, starting from the Gelmini-Roncadelli Majoron model, a rich phenomenology was derived~\cite{Georgi1981}. One of the most interesting consequences considered at that time was the possibility of emitting such a Majoron in double-\Beta decays, giving rise to a new final state in which two electrons and a Majoron (and no neutrinos) are present. The decay rate of this process was calculated, and constraints on the neutrino-Majoron coupling were set using existing limits on the decay rate of \onbb decay~\cite{Georgi1981}. In fact, at that time, the possibility of distinguishing \nnbb, \onbb, or double-\Beta decay with the emission of a Majoron was not conceived, and the latter was regarded only as "an interesting possibility, which may confuse the analysis of double-\Beta decay experiments"~\cite{VERGADOS198296}. 
Only the impressive theoretical work of Doi, Kotani, and Takasugi in 1985, in which the energy and angular distribution of the electrons emitted in the double-\Beta decay with the emission of a Majoron were calculated for the first time~\cite{Doi1985}, provided a clear experimental signature to distinguishing different double-\Beta decay channels. The same authors revised and updated these calculations in a successive work~\cite{Doi1988}. In the same years, a supersymmetric model was developed, which would lead to the emission of two Majorons in double-\Beta decays~\cite{Mohapatra1988}.   

It was clear since early works that the Majoron with nontrivial weak isospin properties (such as the triplet and doublet Majoron), which would have appreciable coupling to neutrinos, would also have a strong coupling to the other leptons and would necessarily be discovered in the upcoming lepton-lepton collisions machines~\cite{Georgi1981}. As the first LEP data on the invisible width of the $Z$ boson became available in the early 90s, the number of active light neutrino species was limited to three, ruling out both triplet and doublet Majoron models~\cite{Steinberger1992, Schael2006}. 

On the other hand, in the same years, the first direct double-\Beta decay experiments started to search for double-\Beta decay with Majoron emission, and an excess of events below \qbb was observed, which could be compatible with the energy distribution predicted by Doi {\it Et al.} for such decay~\cite{Elliott1987, Avignone1991, Fisher1987}. The only available Majoron model not contradicting LEP data was the singlet Majoron model. Nevertheless, in its original Chikashige-Mohapatra-Peccei formulation, the Majoron is so weakly coupled to neutrinos that it cannot produce detectable effects in double-\Beta decay and, therefore, not explain the experimental data.  

The sum of these events motivated in the following years the flourishing of a number of new models able to reconcile the results on the $Z$ decay width with a neutrino-Majoron coupling strong enough to explain the event excess: new models with a singlet Majoron~\cite{Berezhiani1992}, models in which the Majoron carries a non-zero lepton number~\cite{Burgess1993, Burgess1994}, models predicting the emission of two Majorons~\cite{Bamert1995}, and models in which, departing from the original conception of the Majoron as Goldstone boson, the Majoron arises as the component of a massive gauge boson~\cite{Carone1993} or a bulk field~\cite{Mohapatra2000}.

This ensemble of models can lead to two different final states, corresponding to the emission of one or two Majorons, which we will indicate with J:
\begin{equation}
\begin{aligned}
    (A,Z) &\rightarrow (A,Z+2) + 2e^- + J, \\
    (A,Z) &\rightarrow (A,Z+2) + 2e^- + 2J. 
\end{aligned}
\end{equation}

The rate of the double-\Beta decay with the emission of one or two Majorons can be expressed as:
\begin{equation}\label{eq:majoron_conv}
\begin{aligned}
    \relax \Gamma^J &= g_{J\alpha}^2 \, |\mathcal{M}^J_\alpha|^2 \, \mathcal{G}^{J}_\alpha \, , \\
    \Gamma^{JJ} &= g_{J\alpha}^4 \, |\mathcal{M}^{JJ}_\alpha|^2 \, \mathcal{G}^{JJ}_\alpha \, ,
\end{aligned}
\end{equation}
where $g_{J\alpha}$ is the neutrino-Majoron coupling, $\mathcal{M}^{J(JJ)}_\alpha$ the NME, and $\mathcal{G}^{J(JJ)}_\alpha$ the phase-space factor. All three terms depend on the particular model, which we indicated with the subscript $\alpha$. 
Systematic calculations of phase-space factors and NMEs for a number of Majoron models were performed in~\cite{Hirsch1996} for many isotopes. More recently, improved calculations of the summed electron energy distributions have been performed, leading to improved calculations of the phase-space factors~\cite{Kotila2015}. Improved calculations of the NMEs have also been performed lately~\cite{Kotila2021}. 

If one or two Majorons are emitted in the double-\Beta decay, they would escape any detector and carry away part of the decay energy. In analogy with the \nnbb decay, the summed electron energy is continuously distributed between 0 and \qbb, and its exact shape is primarily determined by the phase space. 
In turn, the phase space depends on the Majoron model, particularly on the effective neutrino-Majoron interaction Lagrangian leading to the Majoron-emitting double-\Beta decay.
This can be parameterized to-a-first-approximation with a spectral index $n$:
\begin{equation}
    \frac{d\Gamma^J}{dK} \propto \frac{d\mathcal{G}^J}{dK} \sim (\qbb-K)^n
\end{equation}

\begin{figure}[tbp]
    \includegraphics[width=\textwidth]{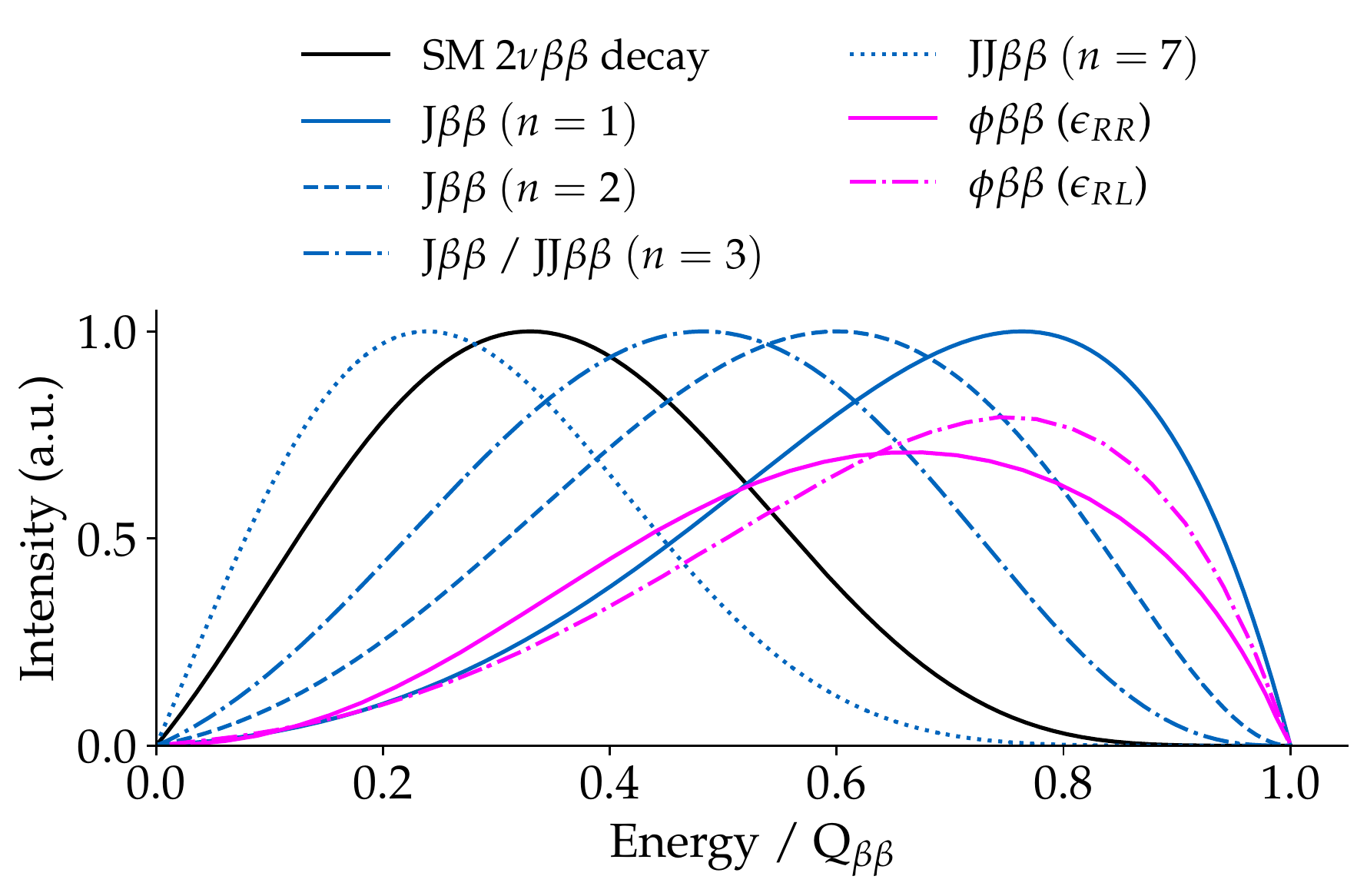}
\caption{Summed electron energy distribution for different Majoron models (the spectral index corresponding to each model is indicated) compared to the SM \nnbb decay distribution. The decays with the emission of a non-standard Majoron are also shown: they can be triggered by an effective seven-dimension operator, containing right-handed ($\epsilon^{\phi}_{RR}$) and left-handed ($\epsilon_{RL}^\phi$) hadronic current. The latter two were adapted from~\cite{Cepedello2019}. An arbitrary normalization is used for illustrative purposes.\label{fig:shape_nnbb_Majorons}}
\end{figure}

Figure~\ref{fig:shape_nnbb_Majorons} shows the summed electron energy distribution for different Majoron models compared to the SM \nnbb decay distribution. 

\begin{table}[tbp]
    \begin{tabular}{ccccc} \toprule
          & Decay & $n$ & Goldstone boson & L \\ \midrule
        \multirow{3}{*}{a} & \multirow{3}{*}{\Jbb} & \multirow{3}{*}{1} & yes & 0 \\
         & & & no & 0 \\
         & & & no & -2\\
         \addlinespace
         b & \Jbb & 2 & Bulk field & 0\\
         \addlinespace
         \multirow{2}{*}{c} & \multirow{2}{*}{\Jbb} & \multirow{2}{*}{3} & yes & -2 \\
         & & & Gauge boson & -2\\
         \addlinespace
         \multirow{3}{*}{d} & \multirow{3}{*}{\JJbb} & \multirow{3}{*}{3} & yes & 0 \\
         & & & no & 0 \\
         & & & no & -1 \\
         \addlinespace
         e & \JJbb & 7 & yes & -1\\ \bottomrule
    \end{tabular}
    \caption{Different Majoron models which predict double-\Beta decays with the emission of one or two Majorons. The third column indicates the model's spectral index ($n$), the fourth column indicates whether the Majoron is a Goldstone boson or not, and the last column indicates the leptonic charge (L) of the Majoron. Models with a leptonic charge different from zero preserve the lepton number symmetry.}
    \label{tab:summary_majorons}
\end{table}

The spectral index is commonly used to group models predicting the same experimental signature, {\it i.e.} the same summed electron energy distribution. These models are not distinguishable by the experiments. Table~\ref{tab:summary_majorons} shows a summary of all the Majoron models grouped by the number of emitted Majorons in the second column, the spectral index in the third column, and the Majoron's properties in the last two columns. The fourth column indicates whether the Majoron is a Goldstone boson or not, whereas the last column shows the Majoron's leptonic charge. Models in which the Majoron carries a lepton number different from 0 preserve the lepton number symmetry. Without an independent test of Lepton number violation, these models are experimentally indistinguishable from the corresponding lepton number non-conserving processes.   

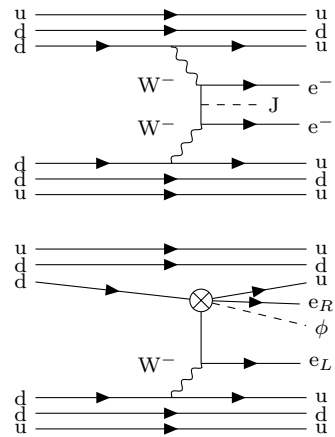
\begin{figure}[tbp]
    \centering 
    \subfloat{
        \centering
        \begin{tikzpicture}
        \begin{feynman}
            \vertex 					(b1) 	{\footnotesize u};
            \vertex[right =4cm of b1] 	(b2) 	{\footnotesize u};
        
            \vertex[below=0.64em of b1] 	(b3) 	{\footnotesize d};
            \vertex[right=4cm of b3] 	(b4) 	{\footnotesize d};
        
            \vertex[below=0.64em of b3] 	(b5) 	{\footnotesize d};
            \vertex[right=2cm of b5] 	(b6);
            \vertex[below=0.64em of b4] 	(b7) 	{\footnotesize u};
        
            \vertex[below=4.8em of b5] 	(c1) 	{\footnotesize d};
            \vertex[right =2cm of c1] (c2);
            \vertex[below=4.8em of b7]	(c3) 	{\footnotesize u};
        
            \vertex[below=0.64em of c1] 	(c4) 	{\footnotesize d};
            \vertex[right=4cm of c4] 	(c5) 	{\footnotesize d};
        
            \vertex[below=0.64em of c4] 	(c6) 	{\footnotesize u};
            \vertex[right=4cm of c6] 	(c7) 	{\footnotesize u};
        
            \vertex[below=1.6em of b7]	(e1)	{\footnotesize e$^-$};
            \vertex[left=1.6cm of e1]		(e2) ;
            \vertex[below=0.8em of e2]  (em1);
            \vertex[right=2.4em of em1]  (em2) {\footnotesize J};
            \vertex[below=1.6em of e2]	(e3) ;
            \vertex[below=1.6em of e1]	(e4)	{\footnotesize e$^-$};
        
            \diagram[small]
            {
            {[edges=fermion]
            (b1) -- (b2),
            (b3) -- (b4),
            (b5) -- (b6),
             (b6) -- (b7),
             (c1) -- (c2),
             (c2) -- (c3),
             (c4) -- (c5),
             (c6) -- (c7),
            (e2) -- (e1),
            (e3) -- (e4),
            },
            (b6) -- [boson, edge label' = W$^-$, font=\footnotesize] (e2),
            (c2) -- [boson, edge label = W$^-$, font=\footnotesize] (e3),
            (e2) -- [plain] (e3),
            (em1) -- [scalar] (em2),
            };
        \end{feynman}
        \end{tikzpicture}
    }
    \\
    \subfloat{
        \centering
        \begin{tikzpicture}
        \begin{feynman}
            \vertex 					(b1) 	{\footnotesize u};
            \vertex[right =4cm of b1] 	(b2) 	{\footnotesize u};
        
            \vertex[below=0.64em of b1] 	(b3) 	{\footnotesize d};
            \vertex[right=4cm of b3] 	(b4) 	{\footnotesize d};
        
            \vertex[below=0.64em of b3] 	(b5) 	{\footnotesize d};
            \vertex[left=1.6cm of b4] (b6);
            \vertex[below=1em of b6, crossed dot] 	(b6c) {};
            \vertex[below=0.64em of b4] 	(b7) 	{\footnotesize u};
        
            \vertex[below=4.8em of b5] 	(c1) 	{\footnotesize d};
            \vertex[right=2cm of c1] (c2);
            \vertex[below=4.8em of b7]	(c3) 	{\footnotesize u};
        
            \vertex[below=0.64em of c1] 	(c4) 	{\footnotesize d};
            \vertex[right=4cm of c4] 	(c5) 	{\footnotesize d};
        
            \vertex[below=0.64em of c4] 	(c6) 	{\footnotesize u};
            \vertex[right=4cm of c6] 	(c7) 	{\footnotesize u};
        
            \vertex[below=1.em of b7]	(e1)	{\footnotesize e$_R$};
            \vertex[left=1.6cm of e1]		(e2) ;
            \vertex[below=2.em of b7] (m1) {\footnotesize $\phi$};
            \vertex[below=2.4em of e2]	(e3) ;
            \vertex[below=2.4em of e1]	(e4)	{\footnotesize e$_L$};
    
        \diagram[small]
        {
        {[edges=fermion]
        (b1) -- (b2),
        (b3) -- (b4),
        (b5) -- (b6c),
         (b6c) -- (b7),
         (c1) -- (c2),
         (c2) -- (c3),
         (c4) -- (c5),
         (c6) -- (c7),
        (b6c) -- (e1),
        (e3) -- (e4),
        },
        (c2) -- [boson, edge label = W$^-$, font=\footnotesize] (e3),
        (b6c) -- [plain] (e3),
        (b6c) -- [scalar] (m1),
        };
        \end{feynman}
        \end{tikzpicture}
    }
\caption{(Top) Feynman diagram for the double-\Beta decay with the emission of a Majoron in classical models. (Bottom) Feynman diagram for the emission of a Majoron-like particle $\phi$ through an effective dimension-seven operator containing right-handed currents in double-\Beta decay (adapted from~\cite{Cepedello2019}). }
\label{fig:diagrams_majo}
\end{figure}

All the models discussed so far focus on the emission of one or two Majorons originating from the intermediate neutrino exchanged in the process. 
Despite the differences among the models, {\it i.e.} the different effective neutrino-Majoron interaction Lagrangians, all of them assume the SM V-A structure of the charged currents involving leptons and quarks. We will refer to them as "classical" models.
Recently a new scenario has been considered, in which a Majoron-like particle ($\phi$) is emitted in the double-\Beta decay (\phibb decay). In this case, the interaction is described by an effective dimension-seven operator, with right-handed lepton current and right/left-handed quark current~\cite{Cepedello2019}. The Feynman diagram of this process is shown in figure~\ref{fig:diagrams_majo} in comparison to the Majoron emission in classical models. The coupling strength between the neutrino and the Majoron-like $\phi$ is $\epsilon_{RL}$ if the effective operator contains left-handed quark current, and $\epsilon_{RR}$ when the effective operator contains right-handed quark current. The two cases have been considered separately, with only one of the two operators being present at a time. 

The energy distribution predicted for the \phibb decay is also shown in figure~\ref{fig:shape_nnbb_Majorons}. The distribution associated with $\epsilon_{RL}$ is very similar to the classical Majoron emission models leading to $n=1$. On the other hand, introducing a hadronic right-handed current in the $\epsilon_{RR}$ term changes the shape of the distribution considerably. 

In all the previous discussions, we always assumed the Majoron to be massless. However, many of the models already presented do not prevent the Majoron from being a light particle~\cite{Bamert1995, Blum2018, Cepedello2019}. This possibility became extremely popular because light Majorons could be a dark matter candidate~\cite{Berezinsky1993, Brune2019}. If the Majoron mass is below the \qbb, double-\Beta decay with the emission of a Majoron can still happen. In this case, the end-point of the energy distribution is shifted to $\qbb - m_J$, where $m_J$ is the Majoron mass.  

\subsubsection{Fermions}\label{subsubsec:fermions}

In many extensions of the Standard Model, new spin $1/2$ particles, singlet under the Standard Model gauge group, are introduced in relation to the question of neutrino mass generation or dark matter. Currently, the most popular exotic fermion is the sterile neutrino $N$. Sterile neutrinos are neutral and right-handed Standard Model singlet fermions that interact with ordinary matter only through mixing with the active neutrinos. We refer the reader to Ref.~\cite{Dasgupta2021} for a recent review of the theoretical and experimental motivation for sterile neutrinos, as well as their phenomenological consequence. In a variant of this scenario, the singlet fermion could be furnished with a \Ztwo symmetry, so it can only be produced in pairs. In general, when a new exotic fermion is introduced in the theory, its mass and coupling to the Standard Model particles are free parameters of the model. It is left to laboratory experiments and astrophysical and cosmological observations to probe the vast allowed parameter space. 

In 1980, Shrock examined the possibility of searching for sterile neutrinos in \Beta decays~\cite{Shrock1980}. The admixture of one or more sterile neutrino states would create a discontinuity in the \Beta decay spectrum similar to the discontinuity that a non-zero neutrino mass is expected to create at the endpoint. The position and amplitude of this kink would give information on the mass of the sterile neutrino and its mixing with the active neutrinos. Since then, several \Beta decay experiments have searched for sterile neutrinos and, still today, they set the most stringent constraints in the mass range between $\sim 10$\,eV and $\sim 1$\,MeV~\cite{Dragoun:2015oja, Riis2011, Mertens2015, Abada2019, Bolton:2019pcu}. 

In analogy with the case of single-\Beta decays, sterile neutrinos with masses below few MeV could be produced in double-\Beta decays. This possibility was recently discussed in~\cite{Bolton2020, Agostini2020}. In~\cite{Agostini2020}, the production of exotic fermions in double-\Beta decays was also extended to models in which the single production is forbidden by additional symmetries while the pair production is allowed. Such a model could be realized with the neutral singlet fermion $\chi$ -- a potential dark matter candidate -- being charged under a discrete \Ztwo symmetry to make it stable. This new fermion could interact with neutrinos through an effective four-fermion scalar interaction of the form $g_\chi \nu \nu \chi \chi$. 
This case is particularly interesting because such a particle cannot be produced in single-\Beta decays, and double-\Beta decays represent a unique discovery opportunity for laboratory experiments.

In general, models predicting the existence of light exotic fermions coupling with the Standard Model neutrinos can lead to two additional double-\Beta decay final states, corresponding to the emission of one or two exotic fermions, which we will indicate with $f$:
\begin{subequations}\label{eq:massive_fermions}
\begin{align}
    (A,Z) &\rightarrow (A,Z+2) + 2e^- + \overline{\nu} + f, \label{eq:one_fermion}\\
    (A,Z) &\rightarrow (A,Z+2) + 2e^- + 2f. \label{eq:two_fermions}
\end{align}
\end{subequations}

Sterile neutrinos can be produced via both~\ref{eq:one_fermion} and~\ref{eq:two_fermions} decay channels. Provided that both decay channels are kinematically allowed -- {\it i.e.} the \qbb value must be larger than the sterile neutrino mass for the ~\ref{eq:one_fermion} decay channel, and twice the sterile neutrino mass for the~\ref{eq:two_fermions} decay channel -- the total double-\Beta decay rate would become an incoherent sum of three channels: 
\begin{equation}\label{eq:decay_rate_nnbb_sterile}
    \Gamma =  \cos^4\theta \; \Gamma_{\nu\nu} + 2\,\cos^2\theta\,\sin^2\theta \; \Gamma_{\nu N} + \sin^4\theta \; \Gamma_{NN}  \;,
\end{equation}
where $\sin^2\theta$ represents the mixing angle between active and sterile neutrinos. The first term accounts for the SM \nnbb decay, the second one for the decay in which one of the two neutrinos is replaced by a sterile neutrino (equation~\ref{eq:one_fermion} with $f=N$), and the last one for the decay into two sterile neutrinos (equation~\ref{eq:two_fermions} with $f=N$). We shall notice here that this last term is strongly suppressed by a factor of $\sin^4\theta$, making it negligible for experimental searches. 

\begin{figure}
    \includegraphics[width=\textwidth]{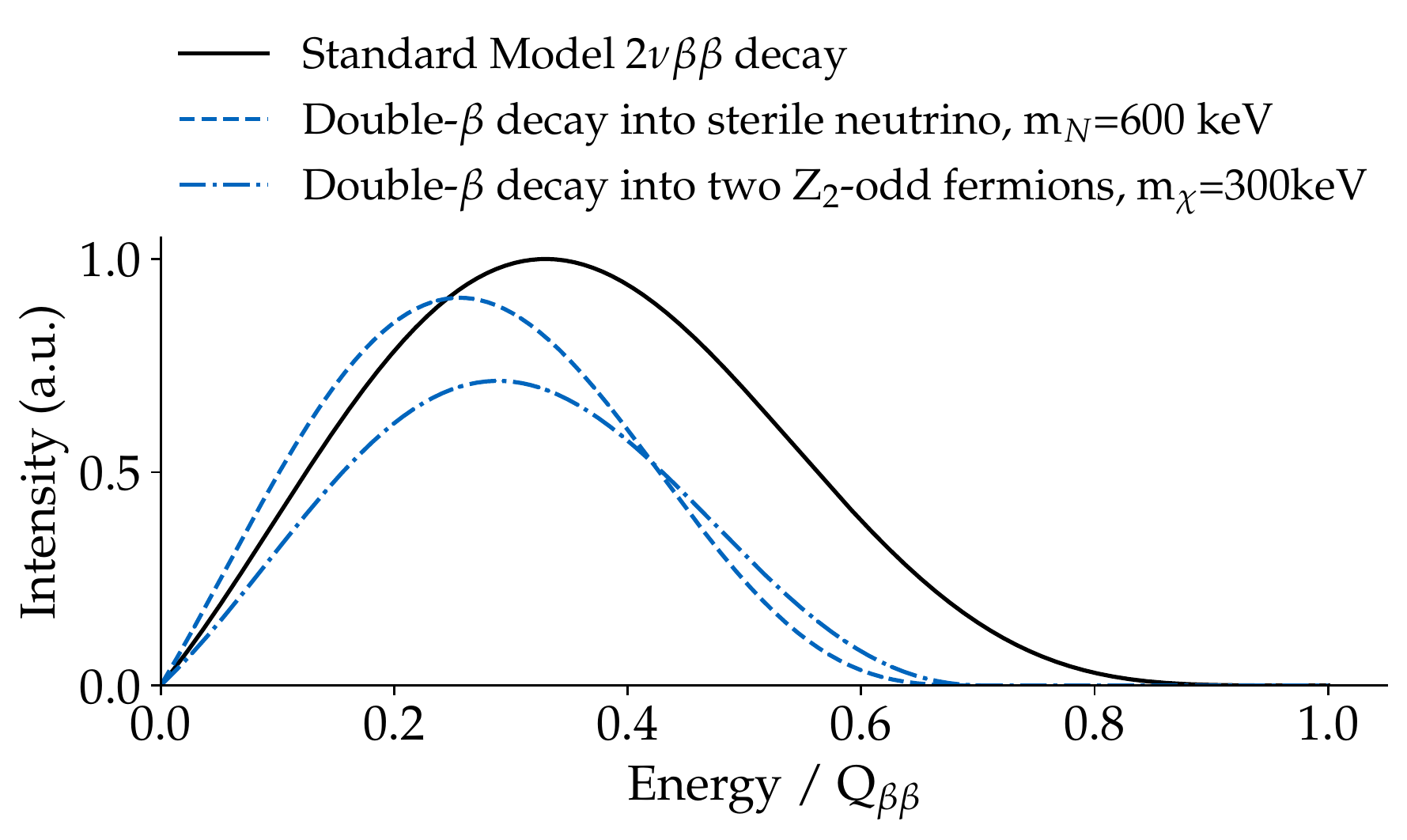}
    \caption{Summed electron energy distributions of the double-\Beta decay into one sterile neutrino with a mass of 600\,keV and the double-\Beta decay into two \Ztwo-odd fermions with a mass of 300\,keV in comparison to the SM \nnbb decay distribution for the case of \Ge isotope. An arbitrary normalization is used for illustrative purposes. Adapted from~\cite{Agostini2020} }
    \label{fig:exotic_fermions_shape}
\end{figure}

Each term of the sum can be expressed factorized as the product of the NME and the phase space factor:
\begin{equation}\label{eq:factorization}
    \Gamma_{ab} = |\mathcal{M}_{2\nu}|^2 \, \mathcal{G}_{ab} \; ,
\end{equation}
where $a$ and $b$ indicate the emitted particles, either the active or the sterile neutrinos.
While the NME is the same for the three terms, the phase space factor is affected by the presence of one or two sterile neutrinos in the final state, and so is the summed electron energy distribution:
\begin{equation}
    \frac{d\Gamma_{ab}}{dK} = |\mathcal{M}_{2\nu}|^2 \, \frac{d\mathcal{G}_{ab}}{dK}
\end{equation}
The presence of a massive sterile neutrino in the final state affects the kinematics of the decay such that the endpoint of the summed electron energy distribution is shifted at $\qbb - m_N$, where $m_N$ is the sterile neutrino mass. The energy distribution for the double-\Beta decay into one sterile neutrino with a mass of 600\,keV is shown in figure~\ref{fig:exotic_fermions_shape} for the case of \Ge isotope.

\Ztwo-odd fermions, which we refer to as $\chi$ fermions, can be produced in double-\Beta decay via the~\ref{eq:two_fermions} channel. To be kinematically allowed, the \qbb value of the decay must be larger than twice the mass of the $\chi$ fermions. The rate of the double-\Beta decay with the emission of two fermions $\chi$ can be expressed as:
\begin{equation}
    \Gamma^{\chi\chi} = \frac{g_\chi^2\, m_e^2}{8 \pi^2 \,R^2} \, |\mathcal{M}^{0\nu}|^2 \, \mathcal{G}^{\chi\chi} \;,
\end{equation}
where $g_\chi$ is the coupling between neutrinos and the $\chi$ fermions, $m_e$ the electron mass, and $R$ the nuclear radius. The NME for this decay can be take to good approximation as the NME for the \onbb decay, $\mathcal{M}_{0\nu}$. The presence of two massive $\chi$ fermions in the final state affects the kinematics of the decay as for the case of sterile neutrinos: the endpoint of the summed electron energy distribution is shifted by $\qbb - 2m_\chi$, where $m_\chi$ is the $\chi$-fermion mass. This is shown for $m_\chi = 300$\,keV in figure~\ref{fig:exotic_fermions_shape} for the case of \Ge isotope.

\subsection{Violation of fundamental symmetries}\label{subsec:violation_symmetry}

\subsubsection{Lorentz violation}

Lorentz invariance is one of the fundamental symmetries of the SM of particle physics. The breakdown of Lorentz and CPT symmetries at the Plank scale is an interesting feature of many theories of quantum gravity, such as string theory~\cite{Kostelecky1989}. Despite direct studies of physics at this ultrahigh energy scale are far from the reach of current accelerator-based experiments, some suppressed effects could arise at lower energies and be  observable with the actual experimental technologies. 

The general framework that characterizes Lorentz violation in the SM is the Standard Model Extension (SME)~\cite{Colladay1997, Colladay1998}. This is an effective quantum field theory that includes all possible operators that can be constructed with the SM fields and that introduce Lorentz violation but preserve the SM gauge invariance. The development of the SME has led to experimental searches for Lorentz violation in all different sectors of physics, including matter, photon, neutrino, and gravity~\cite{Kostelecky2009, Kostelecky2011}. A data table of the current constraints is compiled in~\cite{Kostelecky2011b} annually. 

The behavior of neutrinos in the presence of Lorentz and CPT violation has been extensively studied using the SME framework, and its related coefficients classified~\cite{Kostelecky2004, Kostelecky2004b, Kostelecky2012}. Most of these coefficients are currently constrained by neutrino oscillations experiments. However, four coefficients only affect the neutrino phase space and escape detection through the measurement of neutrino oscillations. The corresponding operators are an example of \emph{counter-shaded} Lorentz-violating operators and they might have escaped detection to date even though their effect is large compared to the suppression given by the Plank scale~\cite{Kostelecky2009}. 
These coefficients, commonly referred to as \emph{oscillation-free} (of) coefficients, can be studied in weak decays, such as single-\Beta decay or double-\Beta decay~\cite{Diaz2013, Diaz2014b}. 

\begin{figure}[tbp]
    \includegraphics[width=\textwidth]{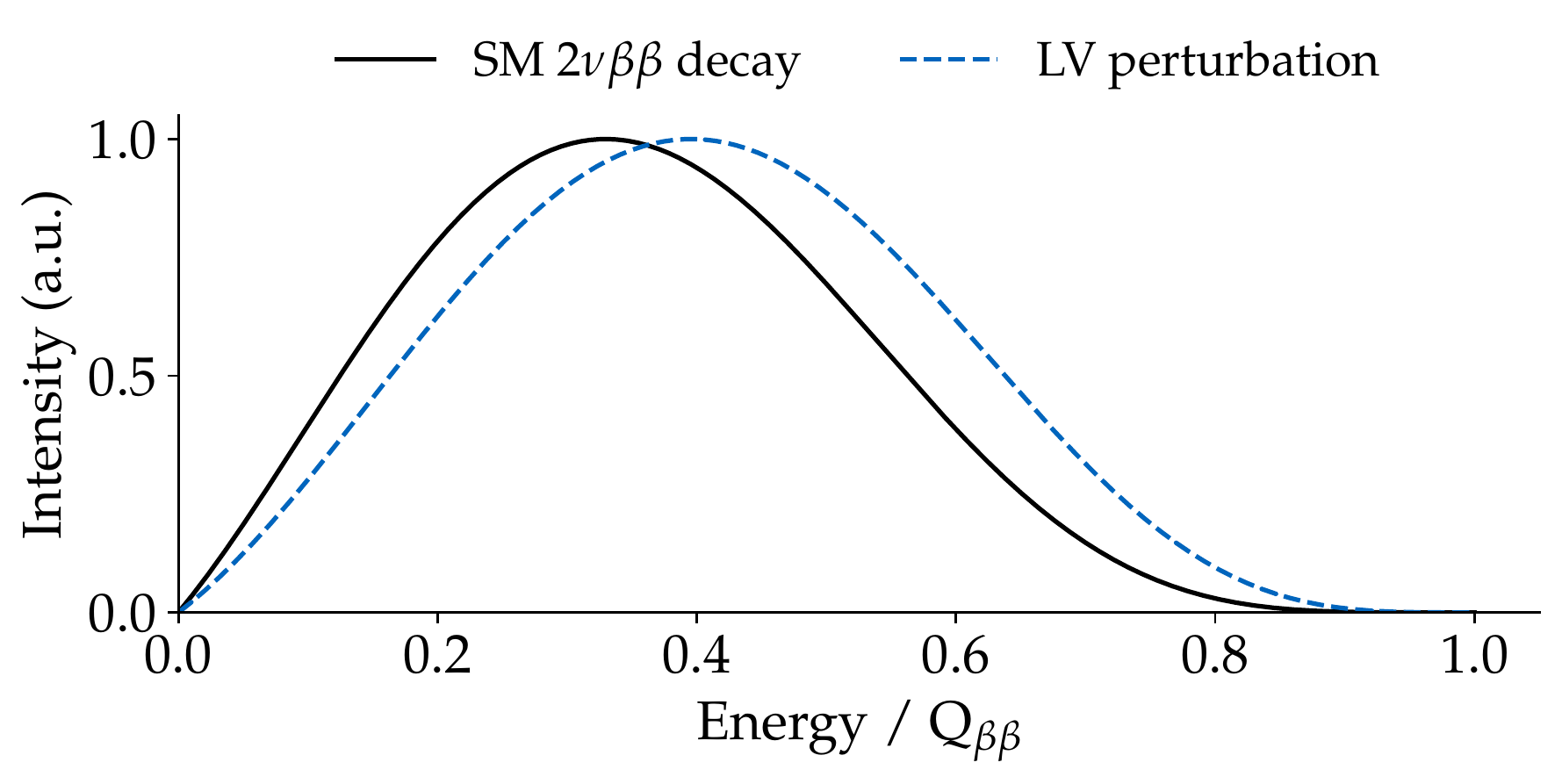}
    \caption{Summed electron energy distribution of the \nnbb decay in the SM and the perturbation term introduced by Lorentz violation (LV). An arbitrary normalization is used for illustrative purposes.}
    \label{fig:shapes_lorentz}
\end{figure}

The interaction of neutrinos with the counter-shaded operator modifies their four-momentum:
\begin{equation}
    p = (E, \mathbf{p}) \longrightarrow p = (E, \mathbf{p} + \mathbf{a}_{of}^{(3)}- \alv \hat{\mathbf{p}}) 
\end{equation}
where $\mathbf{a}_{of}^{(3)}$ encodes the \emph{oscillation-free} coefficients and \alv is the isotropic component $\alv \equiv (a_{of}^{(3)})_{00}/\sqrt{4\pi}$. Considering this modification, the decay rate of the \nnbb decay in the SME framework can be written as the sum of two terms:
\begin{equation}\label{eq:lorentz_violation_decay}
    \Gamma^{2\nu}_{SME} = \Gamma_{SM} + \delta \Gamma_{LV}
\end{equation}
where $\Gamma_{SM}$ is the SM decay rate and $\delta \Gamma_{LV}$ is the perturbation term due to the introduction of Lorentz violation. Similarly to the previous cases, we can factorize each term as the product of the NME and the phase space factor (see equation~\ref{eq:factorization}). Lorentz violation does not affect the NME but appears as a kinematic effect modifying the phase space factor, thus the summed electron energy distribution:
\begin{equation}
    \frac{d\Gamma^{2\nu}_{SME}}{dK} = |\mathcal{M}_{2\nu}|^2 \left( \frac{d\mathcal{G}_{SM}}{dK} + \frac{d(\delta\mathcal{G}_{LV})}{dK} \right)
\end{equation}
The modification of the phase space $d(\delta\mathcal{G}_{LV})/dK$ comes from the change of the differential element of the anti-neutrino momentum:
\begin{equation}
    d^3p = 4\pi\,E^2 dE \longrightarrow d^3p = 4\pi\, (E^2+2E\,\alv)\, dE \;.
\end{equation}
The integration over all anti-neutrino orientation performed to obtain the summed electron energy distribution in the case of double-beta decays implies that only isotropic effects are observable. Hence, the spectrum only depends on \alv.  

The energy dependency in the phase space of the perturbation term can be approximated as $\delta \mathcal{G}_{LV} \sim (\qbb-K)^4$. Using the same terminology introduced for the Majoron, the spectral index of this perturbation is $n=4$. On the other hand, the spectral index of the SM term is $n=5$. Therefore, a non-zero value of the coefficient \alv, which implies a non-zero contribution of the perturbation term, produces a distortion of the spectrum of double-\Beta decays compared to the SM expectation. The energy distribution of the Lorentz violating perturbation term is shown in figure~\ref{fig:shapes_lorentz} compared to the SM \nnbb decay distribution.

\subsubsection{Violation of Pauli exclusion principle}\label{subsec:bosonic_nu}

Neutrinos have many peculiarities among all the known particles. They are the only neutral leptons, which leaves the possibility for neutrinos to be Majorana particles. In addition, the smallness of their mass points to a different mass-generating mechanism compared to the standard coupling with the Higgs boson exhibited by all other particles. Therefore, neutrinos might have substantially different properties compared to the charged leptons. 

Pauli's original formulation of the exclusion principle in 1925 postulated that 
two or more identical electrons cannot occupy the same quantum state within a quantum system simultaneously. This was successively extended to all fermions with half-integer spin  and experimentally confirmed with extremely high precision. From the theoretical point of view, a local quantum field theory with violation of the Pauli principle has been discussed~\cite{Ignatiev1987, Greenberg1988, Okun1987}, and some difficulties highlighted~\cite{Greenberg1989, Govorkov1989}.

The consequences of a change of neutrino statistics from fermionic to bosonic would have substantial cosmological and astrophysical consequences~\cite{CUCURULL1996, Choubey2006, Dolgov2005, Dolgov2005b}. However, even recent analyses of available cosmological data can set only weak bounds on neutrino statistics~\cite{Salas2018}.  With two identical anti-neutrinos in the final state, double-\Beta decay is a unique process to test the violation of the Pauli's principle~\cite{Dolgov2005b, Barabash2007}. Qualitative conclusions in~\cite{Dolgov2005b} on \nnbb decay ruled out a pure bosonic neutrino, but not the possibility that neutrinos obey non-standard statistics, more general than Bose or Fermi ones~\cite{Ignatiev2006}. 

If neutrinos obey a mixed statistic, the neutrino's state would be the combination of fermionic and bosonic states. As a consequence, the double-\Beta decay amplitude would be given by the sum of two terms, corresponding respectively to the fermionic (anti-symmetric) and bosonic (symmetric) parts of the two anti-neutrino emissions:
\begin{equation}
    A_{2\nu\beta\beta} = \cos^2\chi \, A_f + \sin^2\chi \, A_b\;.
\end{equation}
In the phase-space integration, the interference between the anti-symmetric and symmetric parts of the amplitude vanishes and the \nnbb decay rate becomes 
\begin{equation}
    \Gamma_{2\nu\beta\beta} = \cos^4\chi \, \Gamma_f + \sin^4\chi \, \Gamma_b \;,
\end{equation}
where the decay rates $\Gamma_f$ and $\Gamma_b$ are proportional to the squared amplitudes $|A_f|^2$ and $|A_b|^2$, respectively, for pure fermionic and pure bosonic neutrinos. In the decay rates for pure fermionic and pure bosonic neutrinos, both the kinematics terms and the NMEs are different. Defining the ratio
\begin{equation}
    r_0 = \Gamma_b / \Gamma_f \; ,
\end{equation}  
the normalized differential decay rate can be written as
\begin{equation}
    \frac{d\Gamma_{tot}}{\Gamma_{tot}} = \frac{\cos^4\chi}{\cos^4\chi + \sin^4 \chi \, r_0} \frac{d\Gamma_f}{\Gamma_f} + \frac{\sin^4\chi \, r_0}{\cos^4\chi + \sin^4 \chi \, r_0} \frac{d\Gamma_b}{\Gamma_b} \; .
\end{equation}
The ratio $r_0$ determines the weight with which the bosonic component enters the total rate and the differential decay distribution. If $r_0$ is very small, a substantial modification of the energy distribution is expected only for $\sin^2\chi$ being very close to 1. 
In addition, the ratio $r_0$ needs to be calculated and depends on the values of the NMEs. Thus, it introduces an uncertainty due to the nuclear-structure calculations. 

\begin{figure}[tbp]
    \includegraphics[width=\textwidth]{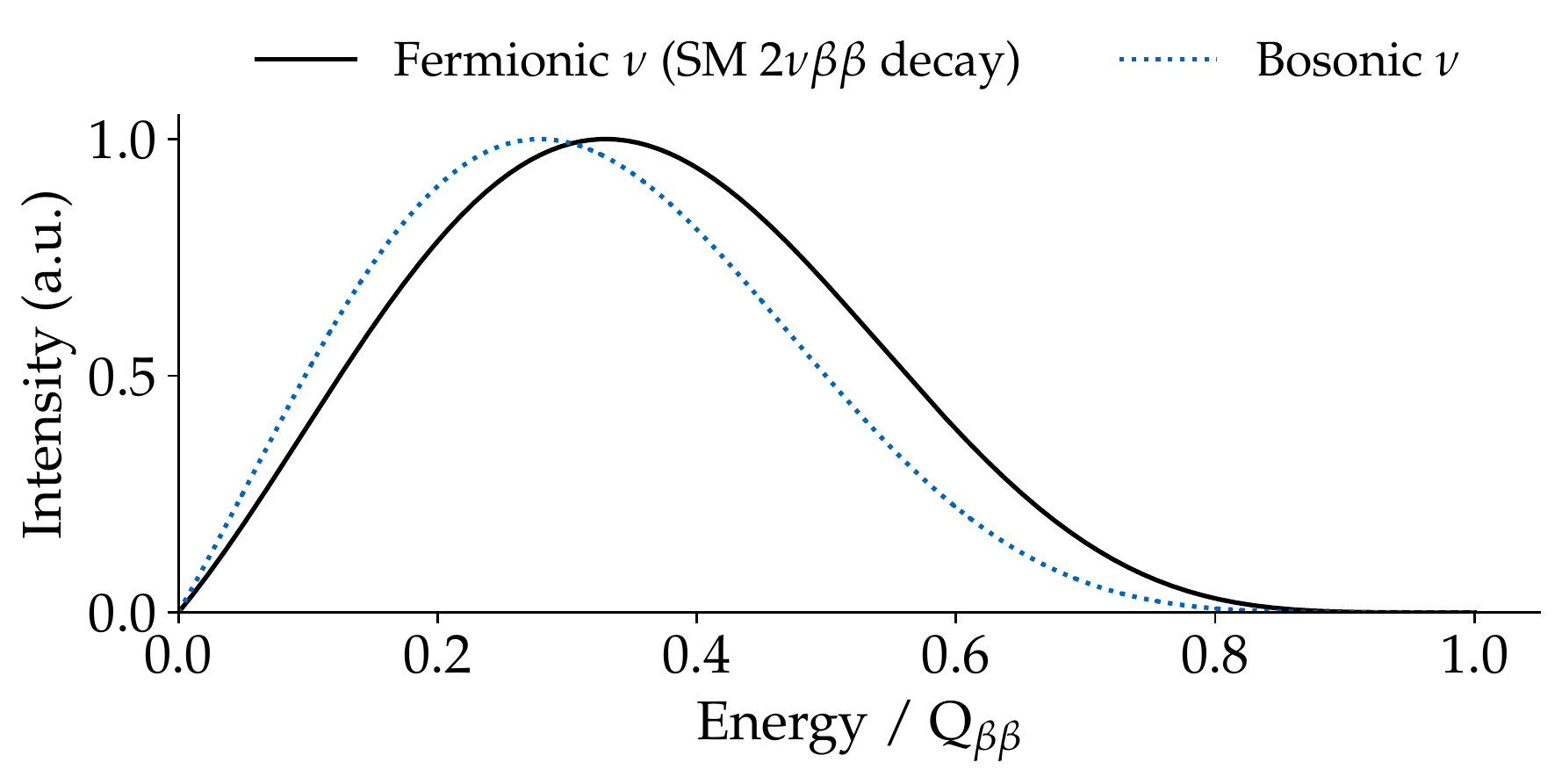}
    \caption{Summed electron energy distribution of the \nnbb decay for pure bosonic neutrinos compared to the case of pure fermionic neutrinos (SM \nnbb decay). An arbitrary normalization is used for illustrative purposes.}
    \label{fig:bosonic_nu_shape}
\end{figure}

On the other hand, the normalized differential decay rate for pure fermionic $d\Gamma_f/\Gamma_f$ and pure bosonic $d\Gamma_b/\Gamma_b$ neutrinos does not depend on any nuclear model assumption and are shown in figure~\ref{fig:bosonic_nu_shape}. The spectrum for bosonic neutrinos is softer, with the maximum shifted at lower energy by a factor of about 15\%, compared to the pure fermionic spectrum. 

The calculations performed in~\cite{Barabash2007} predicts the ratio $r_0$ for \Mo and \Ge to be 0.076 and 0.0014, respectively. The small ratio predicted for \Ge limits the sensitivity of double-\Beta decay experiments with \Ge to spectral distortions due to a partly bosonic neutrino. 

Experimental searches for bosonic or partly bosonic neutrinos with double-\Beta decay experiments could use not only the shape of the distributions but also the ratios between the rates of the transitions to the excited states and the ground states if the first were observed~\cite{Barabash2007}.

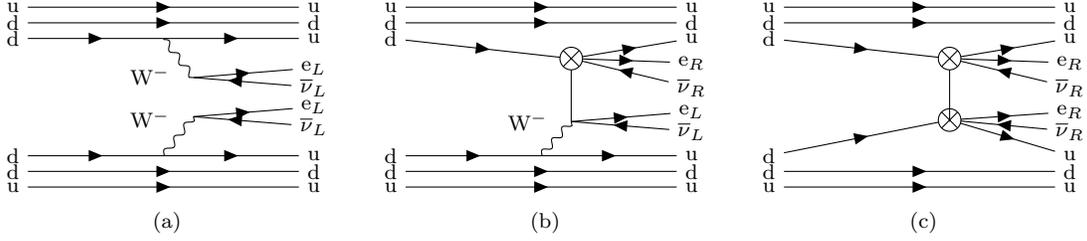
\begin{figure*}
    \centering 
    \subfloat[]{
        \centering
        \begin{tikzpicture}
        \begin{feynman}
    \vertex 					(b1) 	{\footnotesize u};
    \vertex[right =4cm of b1] 	(b2) 	{\footnotesize u};
    
    \vertex[below=0.64em of b1] 	(b3) 	{\footnotesize d};
    \vertex[right=4cm of b3] 	(b4) 	{\footnotesize d};
    
    \vertex[below=0.64em of b3] 	(b5) 	{\footnotesize d};
    \vertex[right=2.cm of b5] 	(b6);
    \vertex[below=0.64em of b4] 	(b7) 	{\footnotesize u};
    
    \vertex[below=4.8em of b5] 	(c1) 	{\footnotesize d};
    \vertex[right =2.cm of c1] (c2);
    \vertex[below=4.8em of b7]	(c3) 	{\footnotesize u};
    
    \vertex[below=0.64em of c1] 	(c4) 	{\footnotesize d};
    \vertex[right=4cm of c4] 	(c5) 	{\footnotesize d};
    
    \vertex[below=0.64em of c4] 	(c6) 	{\footnotesize u};
    \vertex[right=4cm of c6] 	(c7) 	{\footnotesize u};
    
    \vertex[right=0.4cm of b6] (e0);
    \vertex[below=1.6em of e0]   (e1);
    \vertex[below=1.2em of b7] (e2)	{\footnotesize e$_L$};
    \vertex[below=2.em of b7] (e3) {\footnotesize $\overline{\nu}_L$};
    
    \vertex[right=0.4cm of c2] (f0);
    \vertex[above=1.6em of f0]   (f1);
    \vertex[above=2.em of c3] (f2) {\footnotesize e$_L$};
    \vertex[above=1.2em of c3] (f3)	{\footnotesize $\overline{\nu}_L$};
    
    \diagram[small]
    {
    {[edges=fermion]
      (b1) -- (b2),
      (b3) -- (b4),
      (b5) -- (b6),
      (b6) -- (b7),
      (c1) -- (c2),
      (c2) -- (c3),
      (c4) -- (c5),
      (c6) -- (c7),
      (e1) -- (e2),
      (f1) -- (f2),
    },
    {[edges=anti fermion]
    (e1) -- (e3),
    (f1) -- (f3),
    },
    (b6) -- [boson, edge label' = W$^-$, font=\footnotesize] (e1),
    (c2) -- [boson, edge label = W$^-$, font=\footnotesize] (f1),
    };
        \end{feynman}
        \end{tikzpicture}
    \label{fig:sub:diagram_2nbb_first}
    }
    \quad
    \subfloat[]{
        \centering
        \begin{tikzpicture}
        \begin{feynman}
            \vertex 					(b1) 	{\footnotesize u};
            \vertex[right =4cm of b1] 	(b2) 	{\footnotesize u};
        
            \vertex[below=0.64em of b1] 	(b3) 	{\footnotesize d};
            \vertex[right=4cm of b3] 	(b4) 	{\footnotesize d};
        
            \vertex[below=0.64em of b3] 	(b5) 	{\footnotesize d};
            \vertex[left=1.6cm of b4] (b6);
            \vertex[below=1em of b6, crossed dot] 	(b6c) {};
            \vertex[below=0.64em of b4] 	(b7) 	{\footnotesize u};
        
            \vertex[below=4.8em of b5] 	(c1) 	{\footnotesize d};
            \vertex[right=2cm of c1] (c2);
            \vertex[below=4.8em of b7]	(c3) 	{\footnotesize u};
        
            \vertex[below=0.64em of c1] 	(c4) 	{\footnotesize d};
            \vertex[right=4cm of c4] 	(c5) 	{\footnotesize d};
        
            \vertex[below=0.64em of c4] 	(c6) 	{\footnotesize u};
            \vertex[right=4cm of c6] 	(c7) 	{\footnotesize u};
        
            \vertex[below=1.em of b7]	(e1)	{\footnotesize e$_R$};
            \vertex[left=1.6cm of e1]		(e2) ;
            \vertex[below=2.em of b7] (m1) {\footnotesize $\overline{\nu}_R$};
            \vertex[below=2.4em of e2]	(e3) ;
            \vertex[below=2.em of e1]	(e4)	{\footnotesize e$_L$};
            \vertex[below=2.8em of e1] (m2) {\footnotesize $\overline{\nu}_L$};
    
        \diagram[small]
        {
        {[edges=fermion]
        (b1) -- (b2),
        (b3) -- (b4),
        (b5) -- (b6c),
         (b6c) -- (b7),
         (c1) -- (c2),
         (c2) -- (c3),
         (c4) -- (c5),
         (c6) -- (c7),
        (b6c) -- (e1),
        (m1) -- (b6c),
        (e3) -- (e4),
        (m2) -- (e3),
        },
        (c2) -- [boson, edge label = W$^-$, font=\footnotesize] (e3),
        (b6c) -- [plain] (e3),
        };
        \end{feynman}
        \end{tikzpicture}
    \label{fig:sub:diagram_2nbb_second}
    }
    \quad
    \subfloat[]{
        \centering
        \begin{tikzpicture}
        \begin{feynman}
            \vertex 					(b1) 	{\footnotesize u};
            \vertex[right =4cm of b1] 	(b2) 	{\footnotesize u};
        
            \vertex[below=0.64em of b1] 	(b3) 	{\footnotesize d};
            \vertex[right=4cm of b3] 	(b4) 	{\footnotesize d};
        
            \vertex[below=0.64em of b3] 	(b5) 	{\footnotesize d};
            \vertex[left=1.6cm of b4] (b6);
            \vertex[below=1em of b6, crossed dot] 	(b6c) {};
            \vertex[below=0.64em of b4] 	(b7) 	{\footnotesize u};
        
            \vertex[below=4.8em of b5] 	(c1) 	{\footnotesize d};
            \vertex[below=4.8em of b7]	(c3) 	{\footnotesize u};
            \vertex[left=1.6cm of c3] (c2);
            \vertex[above=1em of c2, crossed dot] 	(c2b) {};

            \vertex[below=0.64em of c1] 	(c4) 	{\footnotesize d};
            \vertex[right=4cm of c4] 	(c5) 	{\footnotesize d};
        
            \vertex[below=0.64em of c4] 	(c6) 	{\footnotesize u};
            \vertex[right=4cm of c6] 	(c7) 	{\footnotesize u};
        
            \vertex[below=1.em of b7]	(e1)	{\footnotesize e$_R$};
            \vertex[left=1.6cm of e1]		(e2) ;
            \vertex[below=2.em of b7] (m1) {\footnotesize $\overline{\nu}_R$};
            \vertex[below=2.4em of e2]	(e3) ;
            \vertex[below=2.em of e1]	(e4)	{\footnotesize e$_R$};
            \vertex[below=2.8em of e1] (m2) {\footnotesize $\overline{\nu}_R$};
    
        \diagram[small]
        {
        {[edges=fermion]
        (b1) -- (b2),
        (b3) -- (b4),
        (b5) -- (b6c),
         (b6c) -- (b7),
         (c1) -- (c2b),
         (c2b) -- (c3),
         (c4) -- (c5),
         (c6) -- (c7),
        (b6c) -- (e1),
        (m1) -- (b6c),
        (c2b) -- (e4),
        (m2) -- (c2b),
        },
        (b6c) -- [plain] (e3),
        };
        \end{feynman}
        \end{tikzpicture}
    \label{fig:sub:diagram_2nbb_third}
    }
\caption{Feynman diagrams of the double-\Beta decay (a) with two left-handed currents, \emph{i.e.} the SM \nnbb decay, (b) with one right-handed current, and (c) with two right-handed currents. Adapted from~\cite{Deppisch2020}.}
\label{fig:Feyn_diagram_non-standard_2nbb}
\end{figure*}

\subsection{Non-standard interaction}\label{subsec:non-standard_interactions}
\subsubsection{Right-handed leptonic currrents}

The SM \nnbb decay is a second-order transition involving weak left-handed V-A currents with the strength given by the Fermi constant G$_F$. Some BSM theories, such as Left-Right symmetric models with unbroken lepton number~\cite{Pati1974, Bolton2019}, predict the existence of V+A lepton currents, which can mediate double-\Beta decays~\cite{Deppisch2020}. The new physics effects can be modeled through effective charged current operators containing V+A lepton currents. The strength of these non-standard interactions is given by $\epsilon_{XR}$\,G$_F$, where the small dimensionless coupling $\epsilon_{XR}$ encapsulates the new physics effects. 

Right-handed current interactions are independent of the Majorana or Dirac nature of neutrinos and do not necessarily violate the lepton number. If the neutrino is a Majorana particle, the operators associated with $\epsilon_{LR}$ and $\epsilon_{RR}$ violate the total lepton number by two units and give rise to extra contributions to the \onbb decay~\cite{Doi1983}. In this case, \onbb decay searches set stringent limits of the order $\epsilon_{LR}  \lesssim 3 \times 10^{-9}$, $\epsilon_{RR} \lesssim 6 \times 10^{-7}$ ~\cite{Deppisch2012}. On the other hand, if there exists a sterile neutrino state $\nu_R$ that combines with $\nu_L$ to form a Dirac neutrino, the right-handed current interactions do not necessarily violate lepton number~\cite{Bolton2019}.
The strong theoretical interest is therefore supported by the fact that their observation, along with the non-observation of lepton number violation, would indicate that neutrinos are Dirac fermions~\cite{Bolton2019}. 

\begin{figure}
    \includegraphics[width=\textwidth]{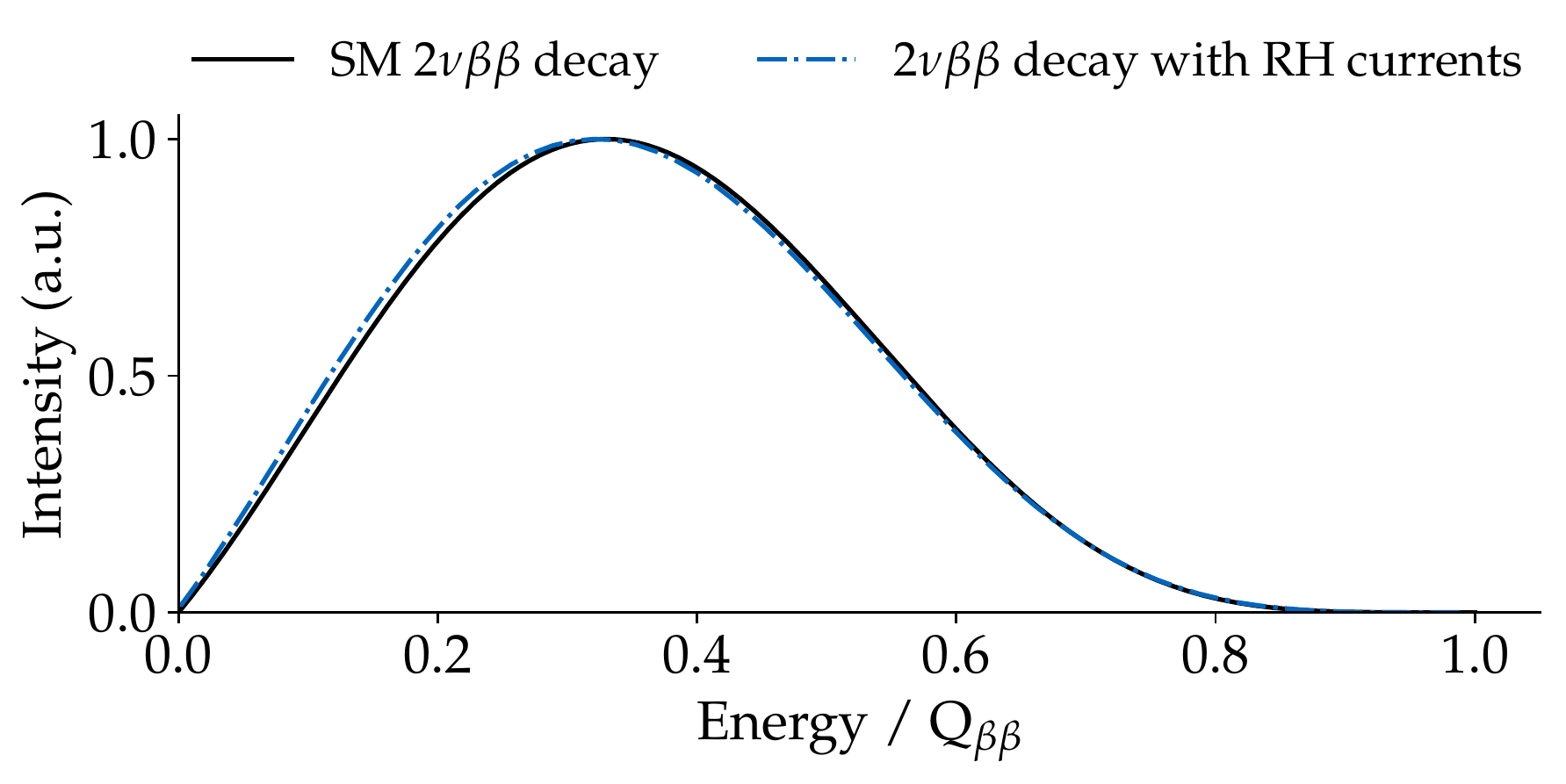}
    \caption{Summed electron energy distribution of the \nnbb decay in the presence of right-handed lepton currents compared to the SM \nnbb decay (left-handed lepton currents). An arbitrary normalization is used for illustrative purposes. Adapted from~\cite{Deppisch2020}.}
    \label{fig:nnbb_shape_right-handed}
\end{figure}

Direct experimental constraints on these operators are set by neutrons and different single-\Beta decays but are rather feeble ($\epsilon_{LR}, \epsilon_{RR}  \lesssim 6 \times 10^{-2}$)~\cite{Gonzalez2019}. Searches at the LHC are also possible~\cite{Greljo2017, Collaboration2014, Collaboration2018} but are generally model dependent and require some caveat on the use of the effective operator analysis at high energies. 

In the presence of right-handed leptonic currents, the amplitude of the \nnbb decay would be calculated as a coherent sum of the three Feynman diagrams shown in figure~\ref{fig:Feyn_diagram_non-standard_2nbb}: the SM second-order transition with two left-handed interactions with the strength given by $\gfermi^2$ (figure~\ref{fig:sub:diagram_2nbb_first}), a transition involving one right-handed interaction with strength $\epsilon_{XR} \, \gfermi^2$ (figure~\ref{fig:sub:diagram_2nbb_second}), and a second-order transition with two right-handed interactions with strength $\epsilon_{XR}^2 \, \gfermi^2$ (figure~\ref{fig:sub:diagram_2nbb_third}). 
Nevertheless, to the lowest order in the exotic coupling $\epsilon_{XR}$, the decay rate can be expressed as an incoherent sum of only two terms:
\begin{equation}
    \Gamma^{2\nu} = \Gamma_{SM} + \epsilon_{XR}^2 \, \Gamma_{\epsilon} \; ,
\end{equation}
where the first term is the SM decay rate and the second term is the contribution of right-handed current to the decay rate, suppressed by the coupling $\epsilon_{XR}$. 
In fact, the interference of the SM term (diagram~\ref{fig:sub:diagram_2nbb_first}) with the diagram~\ref{fig:sub:diagram_2nbb_second} is helicity suppressed by the masses of the emitted electron and neutrino as $m_e m_\nu / \qbb^2$ because of the right-handed nature of the exotic current. Higher orders in the exotic coupling $\epsilon_{XR}$, coming from the last diagram~\ref{fig:sub:diagram_2nbb_third} and its interference with the SM term, are also negligible.     

The phase-space factor and the NME differ in the SM decay rate and the BSM contribution. Thus, the presence of right-handed currents in double-\Beta decay changes the total decay rate and the shape of the energy spectrum. Nevertheless, given the uncertainties in the NME calculations, the change in the total decay rate is not expected to be measurable. Instead, experiments may be sensitive to the change in the spectral shape. Figure~\ref{fig:nnbb_shape_right-handed} shows the \nnbb decay distribution in the SM compared to the distribution arising from the presence of right-handed currents. The deviation includes a spectrum shift to smaller energy and a flatter profile near \qbb.

\subsubsection{Neutrino self-interaction}

The Hubble tension indicates the discrepancy between CMB and local measurement of the Hubble constant. This tension has grown to about $4 \sigma$, and if confirmed, it would require new physics BSM or a new cosmological model~\cite{Riess2019, Aghanim2020}. 

Introducing a \nusi, \emph{i.e.} a four-neutrino contact interaction, could resolve the Hubble tension. Such a \nusi interaction can be written as $G_S(\nu\nu)(\nu\nu)$, and it would inhibit neutrino free-streaming in the early Universe if its strength is much larger than the Fermi effective interaction predicted by the SM, $G_S \sim 10^9$\,\gfermi~\cite{Oldengott2017, Kreisch2020}. This new strong interaction would indicate the presence of new physics at a scale $1/\sqrt{G_S} \sim 10$\,MeV -- 1\,GeV. In general, these strong \nusi interactions are difficult to probe in laboratory experiments due to the absence of electrons or quarks involved. 
With some assumption on the origin of the \nusi operator, constraints can be obtained from different physics observations~\cite{Blinov2019, Lyu2021}, while no model-independent constraint is currently available. The study of \nusi in single-\Beta decays has been considered~\cite{Arcadi2019}. More recently, the search for \nusi in double-\Beta decays has also been proposed~\cite{Deppisch2020b}. 

\begin{figure}
    \centering
    \begin{tikzpicture}
        \begin{feynman}
            \vertex 					(b1) 	{\footnotesize u};
            \vertex[right =4cm of b1] 	(b2) 	{\footnotesize u};
            \vertex[below=0.64em of b1] 	(b3) 	{\footnotesize d};
            \vertex[right=4cm of b3] 	(b4) 	{\footnotesize d};
            \vertex[below=0.64em of b3] 	(b5) 	{\footnotesize d};
            \vertex[right=2cm of b5] 	(b6);
            \vertex[below=0.64em of b4] 	(b7) 	{\footnotesize u};
            \vertex[below=4.8em of b5] 	(c1) 	{\footnotesize d};
            \vertex[right =2cm of c1] (c2);
            \vertex[below=4.8em of b7]	(c3) 	{\footnotesize u};
            \vertex[below=0.64em of c1] 	(c4) 	{\footnotesize d};
            \vertex[right=4cm of c4] 	(c5) 	{\footnotesize d};
            \vertex[below=0.64em of c4] 	(c6) 	{\footnotesize u};
            \vertex[right=4cm of c6] 	(c7) 	{\footnotesize u};
            \vertex[below=1.em of b7]	(e2)	{\footnotesize e$^-$};
            \vertex[above=1.em of c3]	(e3)	{\footnotesize e$^-$};
            \vertex[below=1.8em of b7]	(e1)	{\footnotesize $\nu$};
            \vertex[above=1.8em of c3]	(e4)	{\footnotesize $\nu$};
            \vertex[below=1em of b6] (f1); 
            \vertex[right=1em of f1] (f2);
            \vertex[above=1em of c2] (f3); 
            \vertex[right=1em of f3] (f4);
            \vertex[below=1.5em of f2] (fm);
            \diagram[small]
            {
            {[edges=fermion]
            (b1) -- (b2),
            (b3) -- (b4),
            (b5) -- (b6),
             (b6) -- (b7),
             (c1) -- (c2),
             (c2) -- (c3),
             (c4) -- (c5),
             (c6) -- (c7),
            (fm) -- (e1),
            (fm) -- (e4),
            (f2) -- (fm),
            (fm) -- (f4),
            (f2) -- (e2),
            (f4) -- (e3),
            },
            (b6) -- [boson, edge label' = W$^-$, font=\footnotesize] (f2),
            (c2) -- [boson, edge label = W$^-$, font=\footnotesize] (f4),
            };
        \end{feynman}
    \end{tikzpicture}
    \caption{Feynman diagram of the double-\Beta decay induced by \nusi. Adapted from~\cite{Deppisch2020b}.}
    \label{fig:feynman_nuSI}
\end{figure}
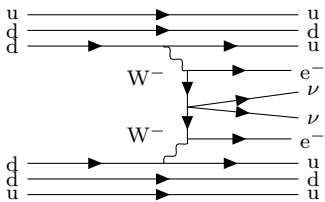

In the presence of \nusi, independently of the Dirac/Majorana nature of neutrinos, the two neutrinos in double-\Beta decay can be emitted via the corresponding effective operator, resulting in a \nusi-induced \nnbb (\nnsibb) decay. The Feynman diagram of this process is shown in figure~\ref{fig:feynman_nuSI}. 
The final state of the \nnsibb decay is identical to that of the SM \nnbb decay. The contribution from \nusi to the decay rate can be written as:
\begin{equation}\label{eq:nuSIdecayrate}
    \Gamma_{\nusi} = \frac{G_S^2 m_e^2}{4R^2} \, \mathcal{G}_{\nusi} \, |\mathcal{M}_{0\nu}|^2 \; ,
\end{equation}
where $m_e$ denotes the electron mass and $R$ the radius of the nucleus. 
For an exact contact interaction of four neutrinos and neglecting the final state lepton momenta, the phase-space factor for the \nnsibb decay is related to the phase-space factor of the \nnbb decay as $\mathcal{G}_{\nusi}=\mathcal{G}_{2\nu}/(4\pi)^2$. The NME of \nnsibb is the same as of \onbb.
In this scenario, no difference is expected in the summed electron energy distribution of the \nnsibb decay compared to the SM \nnbb decay. Therefore, only the experimental measurements of the \nnbb decay rate can be used to constrain the contribution of \nusi. 

\begin{figure}
    \includegraphics[width=\textwidth]{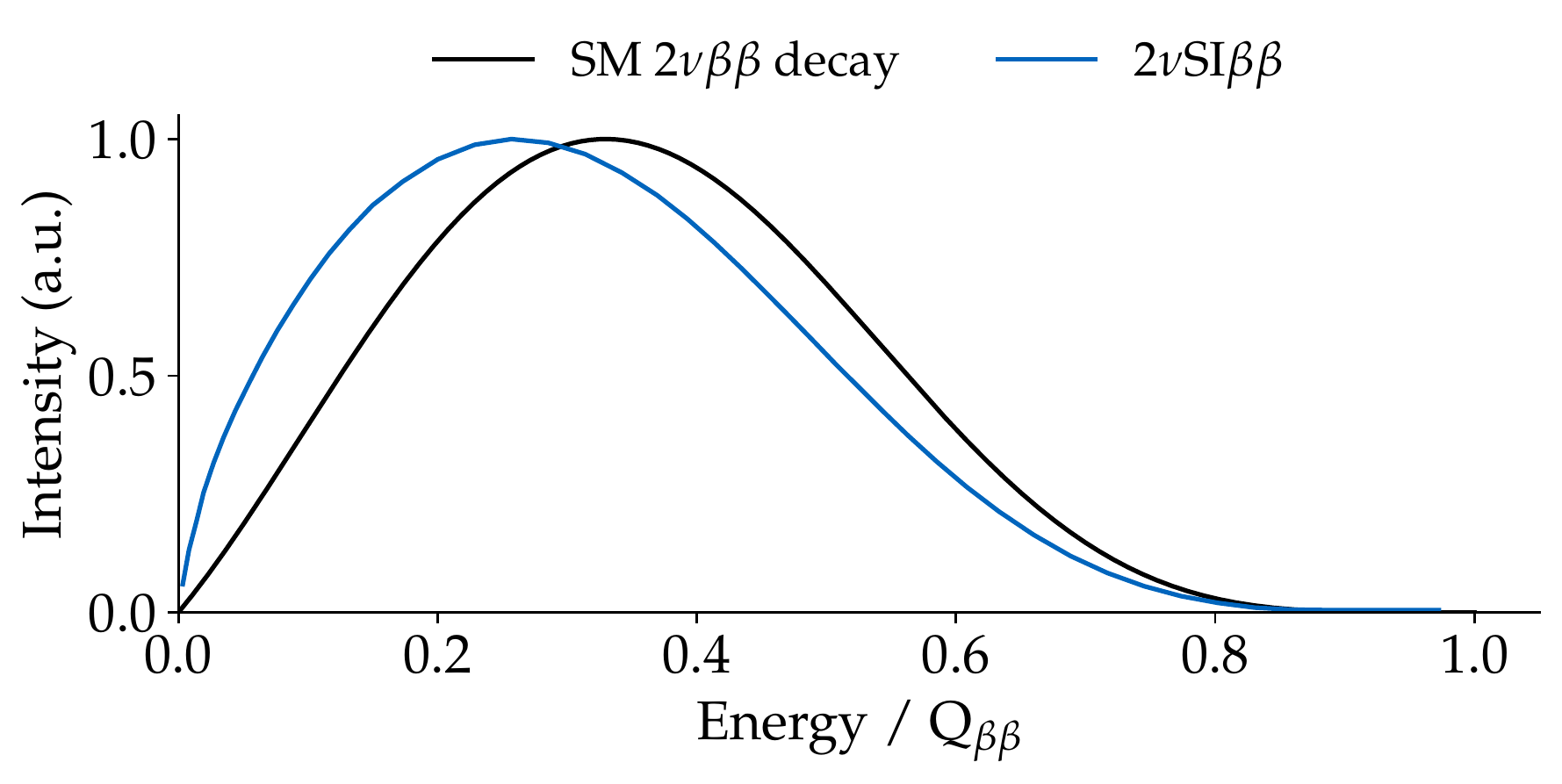}
    \caption{Summed electron energy distribution of \nnsibb decay where the \nusi operator is generated by an s-channel mediator with a mass of $M=\qbb+0.1m_e$, compared to the SM \nnbb decay. An arbitrary normalization is used for illustrative purposes. Adapted from~\cite{Deppisch2020b}.}
    \label{fig:shape_nuSI}
\end{figure}

This approach was used in~\cite{Deppisch2020b} to determine upper limits on the coupling G$_S$ from the measured \nnbb decay rates of several double-\Beta decay isotopes. Limits in the range $G_S/G_F \lesssim (0.32 - 2.50) \times 10^9$ were obtained. The sensitivity on G$_S$ is limited by the uncertainty of the NME ratio $|\mathcal{M}_{0\nu}| / |\mathcal{M}_{2\nu}|$. Cosmological data favoured a strong interactive regime with $G_S = 3.83 \times 10^9 G_F$. Even including the theoretical NME uncertainties, all the considered isotopes can fully exclude the strongly interacting cosmologically favored regime~\cite{Deppisch2020b}. However, one should note that this bound applies only under the assumption that two electron neutrinos are involved in the \nusi. This might not be the case if only muon neutrinos and tau neutrinos participate in \nusi. 

Possible distortions of the electron energy distribution could arise from the \nusi contribution if the \nusi operator were generated by light mediators. In this scenario, the energy dependence of the coupling $G_S$ could cause observable spectral distortions. In~\cite{Deppisch2020b}, the simplest case of an s-channel scalar mediator with a mass just above the kinematic threshold ($M=\qbb+0.1m_e$) was discussed. The coupling $G_S$ acquires the following energy dependence
\begin{equation}\label{eq:gs_nnbb_si}
    G_S = \frac{-M^2}{s-M^2} \, G_S^0 \;,
\end{equation}
where $M$ is the mediator mass and $s \equiv p^2$, with $p$ being the momentum of the mediator (in the context of the \nnsibb, this is of the order $s \lesssim \qbb^2$). The value of $G_S$ at zero momentum transferred ($G_S^0$) is denoted as $G_S^0 = g^2/M^2$, with $g$ the coupling between the mediator and the neutrino. Using $G_S$ in equation~\ref{eq:gs_nnbb_si}, the differential decay rate of the \nnsibb decay can be calculated. The corresponding summed electron energy distribution is shown in figure~\ref{fig:shape_nuSI}, for a mass of the mediator $M=\qbb+0.1m_e$. The energy spectrum of the \nnsibb decay is shifted at lower energy compared to the \nnbb decay spectrum. This shift can be understood qualitatively: with the summed energy of the two electrons increasing, the energy available for the neutrinos is smaller, leading to a smaller value of $s$ and hence a smaller value of $G_S$.

\begin{figure*}
    \centering
    \includegraphics[width=0.8\textwidth]{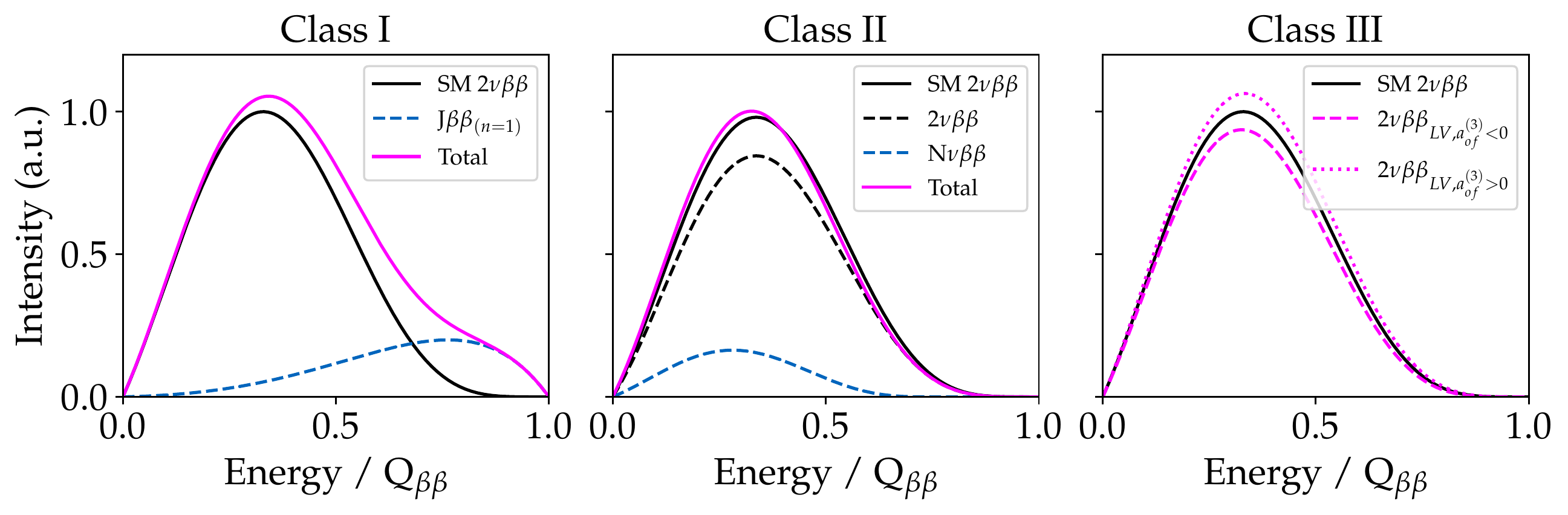}
    \caption{The three classes of BSM double-\Beta decay searches.}
    \label{fig:classes_signatures}
\end{figure*}

\section{Statistical analysis and sensitivity}\label{sec:analysis_methods}

All the double-\Beta decays introduced in section~\ref{sec:models_and_motivations} result in the same event topology. The two electrons emitted in the decay are detected within a small detector region. The additional particles produced in the double-\Beta decay process, either the two anti-neutrinos or one or more exotic particles, escape the detector carrying away part of the decay energy. 

Double-\Beta decay experiments typically measure multiple observables for each event. These include for instance, the energy deposited within the detector, spatial and timing information, and variables related to the type of particles involved in the event. Signal and background events feature specific values of these observables that can be used to separate them. As previously mentioned, this review focuses on the energy deposited within the detector -- the summed energy of the two electrons -- which will be measured by all future \onbb decay experiments. However, we must recall that any additional information, like the energy distribution of the single electrons or their angular correlation, can strongly enhance the sensitivity to BSM physics as it was shown in past searches~\cite{Arnold2016, Arnold2017, Arnold2019}. 

Before discussing the statistical analysis to search for BSM decays, we shall divide them into three classes of models, which require slightly different statistical treatments.  

\paragraph{Class I models.}
The first class contains all those BSM decays that are independent of the SM \nnbb decay. This condition is satisfied if the predicted phase space and NME for the BSM decay are different from the SM ones and if no correlation is expected between the BSM decay and the SM \nnbb decay. This is the case of the decays involving Majorons (\ref{subsubsec:majorons}) or \Ztwo-odd massive fermions (second model in~\ref{subsubsec:fermions}), but also the case of \nnbb decay realized via non-standard interaction (\ref{subsec:non-standard_interactions}), in the assumption that any interference term between the SM \nnbb decay and the non-standard decay can be neglected. Despite the differences among the models, the resulting BSM decay would manifest as an additional contribution to the total energy spectrum, given by a continuous distribution in the same energy range of the SM \nnbb decay. Being the BSM decay subdominant, a small distortion of the observed energy spectrum compared to the SM expectation would be observed. This is shown in the first left panel of figure~\ref{fig:classes_signatures}, for the \Jbb decay (n=1) and a non-realistically large coupling $g_J$, as an example. In the search for such a BSM decay, the parameter of interest is the number of BSM decay events, underlying the respective distribution. On the other hand, the dominant SM \nnbb decay acts as a background. 

\paragraph{Class II models.}
The second class contains those BSM decays which are not independent of the SM \nnbb decay. This is the situation in which the BSM model not only leads to an additional BSM decay but also affects the total decay rate of the SM \nnbb decay. This is the case of the decays involving sterile neutrinos or bosonic neutrinos. The correlation between the SM \nnbb decay and the BSM one is given by the respective mixing, and the existence of the BSM decay also implies a decrease of the SM \nnbb decay by a factor of $cos^4\theta$ and $cos^4\chi$, respectively. This is shown in the central panel of figure~\ref{fig:classes_signatures}, for a sterile neutrino with mass $m_N=500$\,keV and a large mixing of $sin^2\theta=0.15$, as an example. Again, a small distortion of the observed energy spectrum compared to the SM-only expectation would be observed. In the search for such an BSM decay, the parameter of interest is proportional to the ratio between the number of events underlying the BSM decay distribution and the number of events underlying the \nnbb decay distribution. Therefore, in this case, the dominant SM \nnbb decay signal helps to constrain the BSM physics.  
    
\paragraph{Class III models.}    
Some BSM theories do not predict the existence of additional decay, as in class I and class II models, but alter the prediction for the \nnbb decay. We classify these decays as class III. This is the case, for example, of the violation of Lorentz symmetry. The Lorentz violating term in the neutrino momentum affects the phase space of the two electrons emitted in the \nnbb decay and, therefore, the predicted energy distribution. This is shown in the right panel of figure~\ref{fig:classes_signatures}, for non-realistically large values of the Lorentz violating coefficient \alv. The search for Lorentz violation in \nnbb decays corresponds to a search for deviations of the observed two-electron energy distribution compared to the SM expectation. In practice, the total decay rate for the Lorentz violating \nnbb decay can be approximated, to the first order, as a sum of the SM \nnbb decay rate and a perturbation term, whose size is regulated by the coefficient \alv, which is very small. While the phase space differs for the two terms of the sum, the nuclear part, the NME, is the same, introducing a correlation between the two terms. From the statistical point of view, class III models can be treated as class II models. The parameter of interest is proportional to the ratio between the number of events underlying the distribution of the LV perturbation and the number of events underlying the SM \nnbb decay distribution. Again, the dominant SM \nnbb decay signal helps to constrain the BSM physics.

\subsection{Statistical analysis}

In the search for spectral distortions due to the contribution of an BSM decay in the energy spectrum, the energy region of interest extends from the detector threshold to \qbb. In this window, most of the observed events are attributed to the \nnbb decay $N_{2\nu}$ with additional contributions due to different background processes $N_{others}$. The sum of the \nnbb decay and other background constitutes the so-called background model, which is known with a certain accuracy by the experiments. 

To search for an BSM decay, a spectral fit of the energy spectrum is performed,\footnote{Some experiments utilize a multi-variate approach, fitting multiple observable at the same time to better separate signal and background. Here we focus on the simplest approach of one-dimensional fit, but the results can be easily generalized.} adding the BSM decay to the background model. Typically, the information from the background model is used to construct a likelihood function, which is then used in a frequentist or Bayesian approach to constrain the parameter of interest. The definition of the parameter of interest depends on the model to be constrained, as it was pointed out in the previous section.    

The most important experimental parameters determining the sensitivity of the experiment are the exposure $\mathcal{E}$, the background rate $R_{bkg}$, and the systematic uncertainties due to the energy reconstruction $\sigma_{sys}$. The exposure is given by the product of the number of observed nuclei and the observation time. The background rate is primarily given by the \nnbb decay rate with a subdominant contribution due to other sources $R_{bkg}=R_{2\nu}+R_{others}$. The systematic uncertainties related to the energy reconstruction can largely differ between experiments and are specific to the detector technology. 

A precise evaluation of the sensitivity of an experiment requires considering experiment-specific information. However, it can be approximated by considering a counting analysis and Poisson statistics in the region of interest defined above and with a known background expectation given by $R_{bkg}$. In this derivation, we neglect the systematic uncertainties. Their impact will be discussed later in this section. The dependence of the sensitivity on exposure $\mathcal{E}$ and background rate $R_{bkg}$ is different from class I models and class II and II models. In the following, we will discuss the two cases separately. 

\subsection{Experimental sensitivity for class I models}

Let's first consider a BSM decay which we classified as a class I model. This is, for example, the case of the existence of an exotic particle, \emph{e.g.} the Majoron or the \Ztwo-odd exotic fermion, with a coupling to neutrino given by $g_X$.\footnote{The result can be extended to the existence of non-standard interaction, which leads to BSM double-\Beta decays, as introduced in section~\ref{subsec:non-standard_interactions}.} This is the parameter of interest we want to determine the sensitivity. 

The precision with which a subdominant contribution -- the number of BSM decays, $N_X$ -- can be constrained is proportional to the fluctuations of the background in the analysis window, $N_{bkg}$: 
\begin{equation}\label{eq:sensitivity_n}
    N_X \geq \sqrt{N_{bkg}}\;.
\end{equation}
The number of background events in the analysis window can be expressed as a function of the background rate and the exposure:
\begin{equation}\label{eq:nbkg}
    N_{bkg} = R_{bkg} \cdot \mathcal{E} \;,
\end{equation}
and analogously the number of BSM decays:
\begin{equation}\label{eq:nbsm}
    N_X = g_X^2 \, \mathcal{G}\, \mathcal{M}^{2} \cdot \mathcal{E} \;,
\end{equation}
where we have expressed the BSM decay rate as a function of the parameter of interest $g_X$, through the phase-space factor $\mathcal{G}$ and the NME $\mathcal{M}$ of the decay.
Using equations~\ref{eq:nbkg} and~\ref{eq:nbsm} into equation~\ref{eq:sensitivity_n}, we obtain:
\begin{equation}\label{eq:sensitivity_gX}
    g_X^2 \geq \sqrt{\frac{R_{2\nu}+R_{others}}{\mathcal{E}}} \cdot \frac{1}{\mathcal{G}\, \mathcal{M}^{2}} \;.
\end{equation}
In the last step, we have explicated the contribution of the \nnbb decay to the total background rate ($R_{bkg}=R_{2\nu}+R_{others}$). 

The sensitivity scales with the square root of the exposure, but it is limited by the background, to which the \nnbb decay contributes. In addition, equation~\ref{eq:sensitivity_gX} shows that uncertainties in the phase space and NMEs can limit the sensitivity.

\subsection{Experimental sensitivity for class II and III models}

A slightly different result is obtained for class II and III models. Let's consider, for instance, the double-\Beta decay into sterile neutrinos.\footnote{The result can be extended to the double-\Beta decay into bosonic neutrinos and the Lorentz violating \nnbb decay.} In this case, the parameter of interest is the mixing angle \sinT. Given that \sinT also modifies the \nnbb decay rate, it is proportional to the ratio between the number of decays into sterile neutrino $N_{\nu N}$ and the number of \nnbb decay events
\begin{equation}\label{eq:correlation_sinT}
    \sinT \propto \frac{\mathcal{G}_{2\nu}}{\mathcal{G}_{\nu N}} \cdot  \frac{N_{\nu N}}{N_{2\nu}} \;,
\end{equation}
through the respective phase-space factors. 

The statistical uncertainty on this quantity can be computed through standard error propagation
\begin{equation}\label{eq:error_propagation_sinT}
\begin{aligned}
    \frac{\sigma_{\sinT}}{\sinT} \propto 
    \biggl\{ &(\frac{\sigma_{N_{\nu N}}}{N_{\nu N}})^2 + (\frac{\sigma_{N_{2\nu}}}{N_{2\nu}})^2 + \\
    &+2\cdot (\frac{\sigma_{N_{\nu N}}\sigma_{N_{2\nu}}}{N_{\nu N}N_{2\nu}}) \rho_{N_{\nu N},N_{2\nu}} \biggr\}^{1/2} \;.
\end{aligned}
\end{equation}
Because of the same arguments previously used to define the uncertainty on $N_X$, the uncertainty on $N_{\nu N}$ will be 
\begin{equation}
    \sigma_{N_{\nu N}}\propto\sqrt{(R_{2\nu} + R_{others})\cdot\mathcal{E}} \;.
\end{equation}
In addition, we can write $N_{\nu N}$ in terms of $N_{2\nu}$ and \sinT. Also, the correlation coefficient $\rho_{N_{\nu N},N_{2\nu}}$ is proportional to the mixing angle, because of the relation~\ref{eq:correlation_sinT}. Putting everything together, we can rewrite equation~\ref{eq:error_propagation_sinT} as:
\begin{equation}
\begin{aligned}
    \frac{\sigma_{\sinT}}{\sinT} \propto 
    \biggl\{ &\frac{(R_{2\nu} + R_{others})\cdot\mathcal{E}}{R_{2\nu}^2\cdot\sin^4\theta \cdot\mathcal{E}^2} + \dfrac{1}{R_{2\nu}\cdot\mathcal{E}} + \\
    &+2\cdot \frac{\sqrt{R_{2\nu}(R_{2\nu} + R_{others})}\cdot \mathcal{E}}{R_{2\nu}^2\cdot \sin^2\theta \cdot \mathcal{E}^2}
    \biggr\}^{1/2} \;,
\end{aligned}
\end{equation}

\begin{equation}
\begin{aligned}
    \sigma_{\sinT} \propto \biggl\{ &\frac{(R_{2\nu} + R_{others})}{R_{2\nu}^2\cdot\mathcal{E}}+\frac{\sin^4\theta}{R_{2\nu}\cdot \mathcal{E}}+ \\
    &+\frac{2 \cdot \sin^2\theta \sqrt{R_{2\nu}(R_{2\nu} + R_{others})}}{R_{2\nu}^2\cdot \mathcal{E}} \biggr\}^{1/2} \;,
\end{aligned}
\end{equation}

\begin{equation}\label{eq:sensitivity_sinT}
    \sigma_{\sinT} \xrightarrow{\sin^2\theta \sim 0} \mathcal{G}_{2\nu} / \mathcal{G}_{\nu N} \cdot \sqrt{\dfrac{(R_{2\nu}+R_{others})}{R_{2\nu}^2\cdot\mathcal{E}}} \;,
\end{equation}
where we reintroduced the dependence on the phase-space factors in the last passage.

The sensitivity scales with the square root of the exposure, it is limited by the background, but in this case, a high \nnbb decay rate is advantageous due to the dependence $\sigma_{\sinT} \propto \sqrt{1/R_{2\nu}}$. The sensitivity also depends on the ratio between the phase space factors $\mathcal{G}_{2\nu} / \mathcal{G}_{\nu N}$.

\subsection{Impact of the systematic uncertainties on the sensitivity}

\begin{figure}
    \centering
    \includegraphics[width=0.9\textwidth]{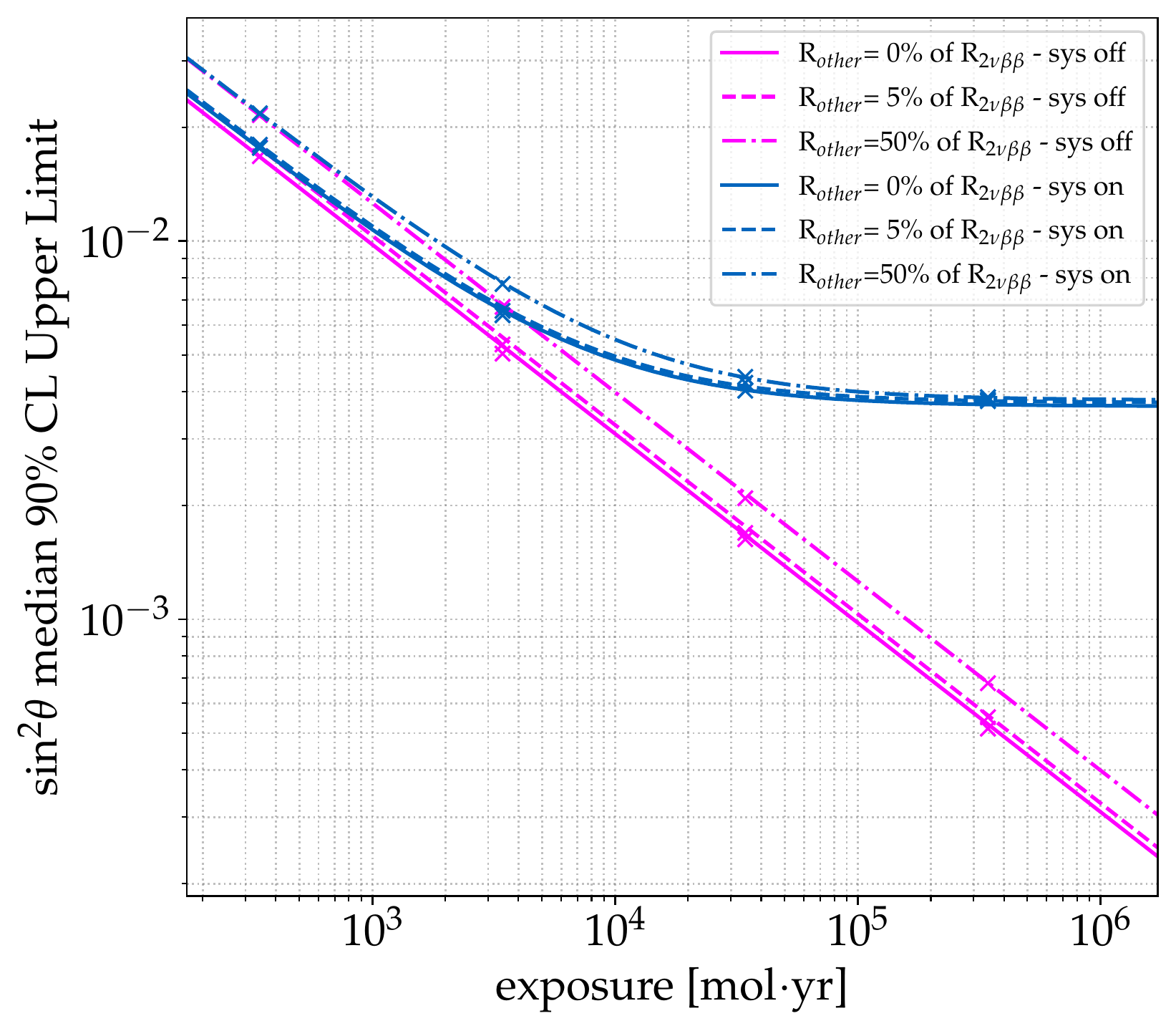}
    \caption{Sensitivity of a double-\Beta decay experiment with \Ge to sterile neutrinos with a mass of 500\,keV, as a function of the exposure, background level, and systematic uncertainties~\cite{Agostini2020}. The results of a full frequentist analysis (markers) are compared to the expectation (lines) from equations~\ref{eq:sensitivity_sys} and~\ref{eq:sensitivity_sinT}.}
    \label{fig:sensitivity_vs_exposure}
\end{figure}

Given an experiment with exposure $\mathcal{E}$ and using an isotope with half-life \bbHL, we expect the sensitivity to double-\Beta decay into Majorons or \Ztwo-odd fermions (or decays resulting from non-standard interaction) to scale with $\sqrt{1/(T_{1/2}^{2\nu} \cdot \mathcal{E})}$ while the sensitivity to the sterile neutrino mixing angle (or Lorentz violation and bosonic neutrinos) with $\sqrt{T_{1/2}^{2\nu}/\mathcal{E}}$. This means that experiments using an isotope with a long \nnbb decay half-life will be favored in the first kind of search as they will have a lower background rate. However, they will be disfavoured in the second kind of search where the sensitivity is linearly proportional to the \nnbb decay half-life.

In our derivation so far, we have neglected the impact of systematic uncertainties on the sensitivity. In a more general way, the sensitivity can be approximated as 
\begin{equation}
    f(\mathcal{E},R_{bkg},\sigma_{sys}) = \sqrt{\sigma_{stat}^2(\mathcal{E},R_{bkg})+\sigma_{sys}^2} \;.
    \label{eq:sensitivity_sys}
\end{equation}
It is determined by both the statistical uncertainty $\sigma_{stat}$, which was derived in equations~\ref{eq:sensitivity_gX} and~\ref{eq:sensitivity_sinT} as a function of the most important experimental parameters, and the systematic uncertainty $\sigma_{sys}$. 

As long as the statistical uncertainty is dominant, the sensitivity improves by increasing the exposure approximately as $f \propto \sqrt{1/\mathcal{E}}$. The sensitivity saturates when the statistical uncertainty becomes comparable with the systematic one. This is illustrated in figure~\ref{fig:sensitivity_vs_exposure}. The figure shows the sensitivity of a double-\Beta decay experiment using \Ge to sterile neutrinos with a mass of 500\,keV, as a function of the exposure, the background level, and the systematic uncertainties~\cite{Agostini2020}. The sensitivity computed using a full frequentist analysis is shown by the markers. This is compared to the expectation given by equation~\ref{eq:sensitivity_sys}, where the statistical uncertainty is given by equation~\ref{eq:sensitivity_sinT}.

\section{Double-\Beta decay experiments and constrains}\label{sec:experimental_results}

The 80-years-long history of double-\Beta decay experiments has seen a variety of technologies and concepts being tested and developed over the years with the standing goal of reducing backgrounds and increasing the isotope mass. A pivotal time for the field was around the turn of the century when the unexpected discovery of neutrino masses~\cite{Super-Kamiokande:1998kpq} raised the question of whether that mass could be due to the peculiar mechanism conceived by Majorana~\cite{Majorana:1937vz, Racah1937, Furry1939}, a hypothesis that can be proven by observing \onbb decay~\cite{Schechter:1981bd}. This boosted the interest for double-\Beta decay experiments, setting in motion a process eventually culminating in the consolidation of five main detection technologies: high-purity germanium semiconductor detectors, cryogenic calorimeters, time projection chambers, large liquid scintillators, and tracking calorimeters.  The following sections review the most recent experiments related to these technologies and the community's plan for the next-generation projects. We conclude with a summary of the state-of-art constraints on the search for BSM double-\Beta decay in Sec.~\ref{sec:constraints} and prospects in Sec.~\ref{sec:prospects_future_searches}.



\subsection{HPGe semiconductor detectors}
High-Purity Ge (HPGe) detectors have been a leading technology for double-\Beta experiments since the very first \onbb decay searches~\cite{Avignone2019}. HPGe detectors are semiconductor devices in which electron-hole charge carriers produced by ionization processes are collected by an electric field applied throughout an ultra-pure Ge crystal isotopically enriched in \Ge up to 92\%. 
Crystals are grown through the Czochralski method~\cite{DEPUYDT2006437} and thus are  intrinsically ultra radio-pure. 
The typical detector size is 1--3\,kg, requiring the simultaneous operation of multiple detectors to reach a large target mass. HPGe detectors have superior energy resolution, the best of any double-\Beta decay experiment, while also providing information on the event topology. Double-\Beta decays are fully contained within the active detector region, and no volume-fiducialization is required to eliminate background leading to very high detection efficiency.  

The most sensitive double-\Beta decay searches based on HPGe detectors have been conducted by the Germanium Detector Array (\gerda) experiment~\cite{Agostini2018} and the \majorana~\cite{Abgrall2014}.  \gerda was located at the Laboratori Nazionali del Gran Sasso (LNGS) in central Italy and operated about 40\,kg of HPGe detectors directly immersed in a LAr volume instrumented to detect its scintillation light.  The \majorana was located in the Sanford Underground Research Facility (SURF) in South Dakota and operated about 30\,kg of HPGe detectors in two vacuum cryostats. The results of the \gerda and \majorana experiments have demonstrated the feasibility of building a ton-scale \Ge-based \onbb decay experiment with ultra-low background and superior energy resolution.  %
With the \gerda and \majorana experiments now completed, the next generation experiment will be realized in the framework of the \legend project, following two stages named LEGEND-200 and LEGEND-1000~\cite{Abgrall2021}. LEGEND-200 has just started approaching physics data taking with 200\,kg of HPGe detectors in the upgraded \gerda infrastructure. LEGEND-1000 is currently under preparation and is expected to come online towards the end of the decade.

The \gerda experiment has performed several searches for BSM double-\Beta decays of \Ge. Its most sensitive search for Majorons-mediated decays led to half-life constraints of $6.4\cdot10^{23}$\,yr, $2.9\cdot10^{23}$\,yr, $1.2\cdot10^{23}$\,yr, and $1.0\cdot10^{23}$\,yr (at 90\% \cl), respectively for the decays with spectral index $n=1,2,3,$ and 7~\cite{Agostini2022}. In the same work, limits on Lorentz violation and the decay into light exotic fermions have also been derived. The Lorentz violating isotropic coefficient \alv has been constrained to $(-2.7<\alv<6.2)\cdot10^{-6}$\,GeV at 90\% \cl. In the search for double-\Beta decays into sterile neutrinos, the most stringent limit was obtained for masses between 500 and 600 keV. For these masses, the 90\% \cl interval obtained on the mixing between sterile and active neutrinos is $\sinT < 0.013$. In the search for double-\Beta decays into \Ztwo-odd fermions, limits on the decay half-life have been derived in the range $(0.18-2.5)\cdot 10^{23}$\,yr, at 90\% \cl. The best limit of $2.5\cdot10^{23}$\,yr was obtained for a mass of 400\,keV and corresponds to a limit on the coupling constant $g_\chi < (0.6-1.4)\cdot 10^{-3}$\,MeV$^{-2}$, where the range is due to the NME uncertainties. 

\subsection{Cryogenic calorimeters}
Cryogenic calorimeters, also referred to as bolometers, have been employed for \onbb decay and dark-matter searches since the 80s~\cite{Brofferio2018}. Bolometers are crystals coupled with a superconductive thermal sensors. Energy deposition within the crystal produce phonons, which warm up the detectors, and additionally scintillation light for specific materials.  
To reach the desire sensitive, the thermal sensors typically rely on the resistivity change of superconductive materials at the transition edge, and thus need to be operated at mK temperatures.

The cryogenic calorimeters are extremely versatile tools, as the crystal material can be chosen in order to study a variety of double-\Beta decay isotopes. Further advantages of this detection technique are an excellent energy resolution and very high detection efficiency. The limited crystals dimensions (typically between 0.2 and 0.8\,kg) require the operation of thousands of cryogenic calorimeters to reach target masses of hundreds of kilograms or larger. This poses challenging requirements to the cryogenic infrastructure that needs to keep a ton of material stably at mK temperatures for years.

Currently, the largest bolometric experiment is the Cryogenic Underground Observatory for Rare Events (CUORE)  at LNGS, which operates about 750\,kg of TeO$_2$ crystals with natural isotopic composition (corresponding to 206\,kg of \Te) in a large cryogen-free cryostat~\cite{Alduino2019}. 
The CUORE experiment successfully demonstrated the feasibility of a ton-scale bolometric experiment~\cite{Adams2022b}, leading to CUPID, the next generation bolometric experiment CUORE Upgrade with Particle Identification enabled by the usage of scintillating bolometers. As part of the R\&D towards CUPID, two independent experiments have produced constraints on BSM physics. The first one is CUPID-0, at LNGS, which utilized ZnSe crystals enriched in \Se~\cite{Azzolini2018}. The second one is CUPID-Mo at LSM in France, which utilized Li$_2$MoO$_4$ crystals enriched in \Mo~\cite{Armengaud2020b}. 

The CUPID-0 experiment has searched for Lorentz violating double-\Beta decay of \Se~\cite{Azzolini2019} and for double-\Beta decays with the emission of Majorons~\cite{CUPID-0:2022yws}. In the first work, the Lorentz violating coefficient \alv has been constrained to $< 4.1 \cdot10^{-6}\,$GeV at 90\% \ci. In the second work, limits on the half-life of the decays involving Majorons have been derived. These are $1.2\cdot10^{23}$\,yr, $3.8\cdot10^{22}$\,yr, $1.4\cdot10^{22}$\,yr, and $2.2\cdot10^{21}$\,yr (at 90\% \ci), respectively for the decays with spectral index $n=1,2,3,$ and 7.

\subsection{Time projection chambers}
The first direct observation of \nnbb decay was made using a Time Projection Chamber (TPC)~\cite{Elliott1987b}. Since then, this technology has been at the forefront of \onbb decay searches because of the combination of mass scalability and optimal background discrimination capabilities enabled by the 3D reconstruction of the event topology, position, and energy. TPCs are particularly well-suited to search for \onbb decay of \Xe: Xe is a noble element that can be used directly in TPCs as a liquid or gas. On the other hand, TPC detectors have a limited energy resolution and require a multi-variate analysis to constrain background features close to the \qbb, which are often not resolved (\emph{e.g.} $^{214}$Bi $\gamma$ line at 2447.7\,keV, just below the \Xe \qbb). 

The most sensitive liquid Xe TPC among the current generation of double-\Beta decay experiments was EXO-200 at WIPP near Carlsbad New Mexico. EXO-200 was a single-phase liquid-Xe TPC, filled with 161\,kg of \Xe~\cite{Auger2012}. EXO-200 demonstrated the capabilities of a monolithic liquid-Xe TPC, which includes relatively good energy resolution, near maximal signal detection efficiency, and solid topological discrimination of backgrounds~\cite{Anton2019}.  Building on it, the next generation ton-scale \Xe-based \onbb decay experiment nEXO is currently being proposed~\cite{Adhikari2022}. 

High-pressure gaseous Xe TPCs for \nnbb-decay have been developed in the context of the NEXT project, which combines electroluminescence to reach good energy resolution ($<1$\% FWHM at \qbb) and charged-particle tracking for the active suppression of background~\cite{Cadenas2019}. 
A concept for a ton-scale phase of NEXT is currently under investigation with the goal of operating a full ton of \Xe in the form of enriched Xe gas~\cite{Adams2021c}, and possibly deploying technologies for the identification of the \Xe decay daughter isotope. 

The EXO-200 experiment has searched for BSM double-\Beta decays of \Xe. It sets upper limits on Majorons emitting decays at the level of 
$4.3\cdot10^{24}$\,yr, $1.5\cdot 10^{24}$\,yr, $6.3\cdot 10^{23}$\,yr, and $5.1\cdot 10^{22}$\,yr  (at 90\% \cl), respectively for the decays with spectral index $n=1,2,3,$ and 7~\cite{Kharusi2021}.
In the same work, limits on the emission of a Majoron-like particle via non-standard RH interaction have also been derived. The half-life of these decays has been constrained at 90\% \cl to $3.7\cdot 10^{24}$\,yr and $4.1\cdot 10^{24}$\,yr, respectively, for right-handed and left-handed quark currents. The EXO-200 experiment first searched for Lorentz violation in \nnbb decay~\cite{Albert2016}. The Lorentz violating coefficient \alv was constrained to $-2.65\cdot 10^{-5}\,\text{GeV} < \alv < 7.60\cdot 10^{-6}\,\text{GeV}$ at 90\% \cl.

\subsection{Large liquid scintillators}
Large liquid scintillator detectors have historically been the most mass-scalable technology used by \onbb experiments. The scintillator can be doped with different isotopes of interest for these searches, including \Xe and \Te. Decays occurring within the detector create scintillation photons, which travel straight to the outer surface of the scintillator volume where they are read out by photo-multipliers tubes. The event position and energy release are reconstructed using the time-of-flight of the photons and their number. Multivariate analysis can also provide some information on the initial spatial extension of the energy deposition, giving an extra handle to tag background-like events.

The most important liquid scintillator detector in the field is operated by the KamLAND-Zen experiment in the Kamioka Mine in Japan~\cite{Albert2017}. It consists of a nylon balloon placed at the center of the detector volume and filled with a liquid scintillator in which \Xe has been dissolved. After a successful first phase (KamLAND-Zen 400) with up to 340\,kg of \Xe, the second phase (KamLAND-Zen 800) is currently running with about 680\,kg of \Xe~\cite{Abe2022}. With the first 1.6 years of data, KamLAND-Zen 800 produced a world-leading \onbb decay half-life limit~\cite{Abe2022}. The KamLAND-Zen collaboration is already preparing for the ton-scale phase (KamLAND2-Zen) in which about 1 ton of \Xe will be deployed~\cite{Shirai:2017jyz}.


The technology of large liquid scintillators is largely employed in neutrino experiments. Among future large liquid scintillator experiments, we shall mention the SNO experiment. This is a multi-purpose neutrino experiment located at SNOLAB in Canada~\cite{Albanese2021}. An upgrade of the experiment is planned after the completion of the main goals, which consists in loading the 780 tons of organic liquid scintillator with a double-beta decaying isotope to search for \onbb decay~\cite{Albanese2021}.


The KamLAND-Zen experiment searched for double-\Beta decays with the emission of Majorons~\cite{KamLAND-Zen:2012uen}. Limits on the half-life of these decays have been derived at 90\% \cl: $2.6\cdot10^{24}$\,yr, $1.0\cdot10^{24}$\,yr, $4.5\cdot10^{23}$\,yr, and $1.1\cdot10^{22}$\,yr, respectively for the decays with spectral index $n=1,2,3,$ and 7.
 
\subsection{Tracking calorimeters}
Tracking calorimeters are the only technology capable of measuring with high accuracy the kinematic of electrons emitted in double-\Beta decays, such as the single-electron energy and the electron angular distribution. Measuring these quantities would give precious inputs to pin down the actual channel mediating the \onbb decay~\cite{Graf:2022lhj}. It would also strongly enhance the sensitivity to search for the BSM double-\Beta decays discussed in this review.

The tracking capability is obtained by decoupling the double-\Beta decay isotope from the detector. The target isotope is placed on a thin foil, immersed in a magnetic field, and surrounded by tracking and calorimetric layers. This configuration enables the measurement of the electron momentum through its bending in the magnetic field, and the measure of its energy when it enters the calorimeters. Unfortunately, it reduces the detection efficiency. The requirement of using very thin foils to minimize energy losses makes it extremely challenging to scale up the isotope mass.

The NEMO-3 experiment utilized this technology to search for \onbb decay of several isotopes at the LSM in France. Masses from a few grams to a few kilograms of the isotopes of interest were deployed in separate sectors of the detector (6.99\,g of \Ca~\cite{Arnold2016b}, 0.932\,kg of \Se~\cite{Arnold2018}, 9.4\,g of \Zr~\cite{NEMO-3:2009fxe}, 6.914\,kg of \Mo~\cite{Arnold2015}, 410\,g of \Cd~\cite{Arnold2017}, and 36.6\,g of \Nd~\cite{Arnold2016}). 
A next-generation tracking calorimeter detector is the SuperNEMO Demonstrator, which is based on the technology demonstrated by NEMO-3~\cite{Piquemal2006}. In its first phase, the SuperNEMO Demonstrator will deploy one module with 7\,kg of \Se.  A future full-scale experiment is foreseen, consisting of multiple modules aiming for a total \Se mass of 100\,kg. 

The NEMO-3 experiment searched for Majoron-involving double-\Beta decays in several isotopes: \Mo~\cite{Arnold2015,Arnold2019}, \Se~\cite{Arnold2018}, \Cd~\cite{Arnold2017}, \Ca~\cite{Arnold2016b} and \Nd~\cite{Arnold2016}. Limits on the half-life of the decay corresponding to spectral index $n=1$ have been derived (at 90\% \cl) for all the used isotopes: $4.6\cdot 10^{21}$\,yr with \Ca, $3.7\cdot 10^{22}$\,yr with \Se, $4.4 \cdot 10^{22}$\,yr with \Mo, $8.5 \cdot 10^{21}$\,yr with \Cd, and $0.3 \cdot 10^{22}$\,yr with \Nd. 
Limits on the half-life of the decays corresponding to spectral indexes $n=2,3,$ and 7 have been derived only with \Mo, because of the lower statistics data sets and the higher background achieved with the other isotopes. The corresponding limits at 90\% \cl are: $9.9 \cdot 10^{21}$\,yr, $4.4 \cdot 10^{21}$\,yr and $1.2 \cdot 10^{21}$\,yr, respectively for the $n=2,3,$ and 7.
The NEMO-3 experiment also searched for Lorentz violation in the \nnbb decay of \Mo and the \Mo \nnbb decay with bosonic neutrinos~\cite{Arnold2019}. The Lorentz violating isotropic coefficient \alv has been constrained to $(-4.2 < \alv < 3.5) \cdot 10^{-7}$\,GeV at 90\% \cl. The half-life of the \Mo \nnbb decay with bosonic neutrinos was constrained to $1.2 \cdot 10^{21}$\,yr, which corresponds to an upper limit on the bosonic neutrino contribution of sin$\chi< 0.27$ at 90\% \cl. 


\subsection{Other technologies}
We should mention the Aurora experiment at LNGS among the experiments using technologies other than those described in the previous subsections. 
It utilized more than 1 kg of radio-pure cadmium tungstate ($^{116}$CdWO$_4$) scintillating crystals enriched in the \Cd isotope~\cite{Barabash:2011hc}. Even if this technology is not competitive in terms of \onbb decay sensitivities and there is no concrete plan to scale it to a ton-scale \onbb decay experiments, the Aurora experiment has produced competitive constraints in the search for BSM double-\Beta decays of \Cd~\cite{Barabash2018}. They set limits on the half-life of Majorons emitting decays with the emission of Majorons at the order of $10^{21}$\,yr, while the most stringent limit was obtained for the $n=1$ spectral index: $T_{1/2} > 8.2\cdot 10^{21}$\,yr (at 90\% \cl). In the same work, they searched for Lorentz violation and obtained a limit on the isotropic coefficient of $\alv < 4.0\cdot 10^{-6}$\,GeV (at 90\% \cl).

\subsection{Most sensitive constraints}\label{sec:constraints}
In this section we summarise the most sensitive constraints reported by all experiments mentioned in the previous sections, grouping them based on the new physics searched.


\paragraph{Double-\Beta decay with the emission of Majorons}

\begin{table*}
    \centering
    \begin{tabular}{ccccccc}
    \toprule
        Decay / Isotope & $T_{1/2}$ (yr) & Experiment & Ref. & G ($10^-{18}$\,yr$^{-1}$) & NME & $g_J$ \\
        \midrule
        \Jbb ($n=1$) \\ \addlinespace
        \Ca & $> 4.6\times 10^{21}$ & NEMO-3      & \cite{Arnold2016b}     & 1540 & $(0.40-2.71)$  & $< (8.5 - 58)\times10^{-5}$  \\
        \Ge & $> 6.4\times 10^{23}$ & GERDA       & \cite{Agostini2022}    & 44.2 & $(2.66-6.64)$  & $< (1.8 - 4.4)\times10^{-5}$ \\
        \Se & $> 3.7\times 10^{22}$ & NEMO-3      & \cite{Arnold2018}      & 361  & $(2.72-5.30)$  & $< (3.2 - 6.2)\times10^{-5}$ \\
        \Se & $> 1.2\times 10^{23}$ & CUPID-0     & \cite{CUPID-0:2022yws} & 361  & $(2.72-5.30)$  & $< (1.8 - 3.5)\times10^{-5}$ \\
        \Mo & $> 4.4\times 10^{22}$ & NEMO-3      & \cite{Arnold2015}      & 598  & $(3.84-6.59)$  & $< (1.8 - 3.1)\times10^{-5}$ \\
        \Cd & $> 8.2\times 10^{21}$ & Aurora      & \cite{Barabash2018}    & 569  & $(3.105-5.43)$ & $< (5.3 - 9.2)\times10^{-5}$ \\
        \Cd & $> 8.5\times 10^{21}$ & NEMO-3      & \cite{Arnold2017}      & 569  & $(3.105-5.43)$ & $< (5.2 - 9.0)\times10^{-5}$ \\
        \Xe & $> 2.6\times 10^{24}$ & KamLAND-Zen & \cite{Gando2012}       & 409  & $(1.11-4.77)$  & $< (0.4 - 1.7)\times10^{-5}$ \\
        \Xe & $> 4.3\times 10^{24}$ & EXO-200     & \cite{Kharusi2021}     & 409  & $(1.11-4.77)$  & $< (0.3 - 1.3)\times10^{-5}$ \\
        \Nd & $> 0.3\times 10^{22}$ & NEMO-3      & \cite{Arnold2017}      & 3100 & $(1.707-5.46)$ & $< (3.7 - 12)\times10^{-5}$  \\
        
        \Jbb ($n=2$)  \\ \addlinespace
        \Ge & $> 2.9\times 10^{23}$ & GERDA       & \cite{Agostini2022}    & -- & -- & -- \\
        \Se & $> 3.8\times 10^{22}$ & CUPID-0     & \cite{CUPID-0:2022yws} & -- & -- & -- \\
        \Mo & $> 9.9\times 10^{21}$ & NEMO-3      & \cite{Arnold2019}      & -- & -- & -- \\
        \Cd & $> 4.1\times 10^{21}$ & Aurora      & \cite{Barabash2018}    & -- & -- & -- \\
        \Xe & $> 1.0\times 10^{24}$ & KamLAND-Zen & \cite{Gando2012}       & -- & -- & -- \\
        \Xe & $> 9.8\times 10^{23}$ & EXO-200     & \cite{Kharusi2021}     & -- & -- & -- \\
        
        \Jbb ($n=3$)  \\ \addlinespace 
        \Ge & $> 1.2\times 10^{23}$ & GERDA       & \cite{Agostini2022}    & 0.073 & 0.381 & $< 1.7\times 10^{-2}$  \\
        \Se & $> 1.4\times 10^{22}$ & CUPID-0     & \cite{CUPID-0:2022yws} & 1.22  & 0.305 & $< 1.5\times 10^{-2}$  \\
        \Mo & $> 4.4\times 10^{21}$ & NEMO-3      & \cite{Arnold2019}      & 2.42  & 0.263 & $< 2.3\times 10^{-2}$  \\
        \Cd & $> 2.6\times 10^{21}$ & Aurora      & \cite{Barabash2018}    & 2.28  & 0.144 & $< 5.6\times 10^{-2}$  \\
        \Xe & $> 4.5\times 10^{23}$ & KamLAND-Zen & \cite{Gando2012}       & 1.47  & 0.160 & $< 0.47\times 10^{-2}$ \\
        \Xe & $> 6.3\times 10^{23}$ & EXO-200     & \cite{Kharusi2021}     & 1.47  & 0.160 & $< 0.40\times 10^{-2}$ \\
        
        \JJbb ($n=3$) \\ \addlinespace  
        \Ge & $> 1.2\times 10^{23}$ & GERDA       & \cite{Agostini2022}    & 0.22 & 0.0026   & $< 1.21$ \\
        \Se & $> 1.4\times 10^{22}$ & CUPID-0     & \cite{CUPID-0:2022yws} & 3.54 & 0.0020   & $< 1.18$ \\
        \Mo & $> 4.4\times 10^{21}$ & NEMO-3      & \cite{Arnold2019}      & 6.15 & 0.0019   & $< 1.41$ \\
        \Cd & $> 2.6\times 10^{21}$ & Aurora      & \cite{Barabash2018}    & 5.23 & 0.000945 & $< 2.37$ \\
        \Xe & $> 4.5\times 10^{23}$ & KamLAND-Zen & \cite{Gando2012}       & 3.05 & 0.0011   & $< 0.69$ \\
        \Xe & $> 6.3\times 10^{23}$ & EXO-200     & \cite{Kharusi2021}     & 3.05 & 0.0011   & $< 0.64$ \\
        
        \JJbb ($n=7$) \\ \addlinespace 
        \Ge & $> 1.0\times 10^{23}$ & GERDA       & \cite{Agostini2022}    & 0.42 & 0.0026   & $< 1.08$ \\
        \Se & $> 2.2\times 10^{21}$ & CUPID-0     & \cite{CUPID-0:2022yws} & 26.9 & 0.0020   & $< 1.13$ \\
        \Mo & $> 1.2\times 10^{21}$ & NEMO-3      & \cite{Arnold2019}      & 50.8 & 0.0019   & $< 1.41$ \\
        \Cd & $> 8.9\times 10^{20}$ & Aurora      & \cite{Barabash2018}    & 33.9 & 0.000945 & $< 2.37$ \\
        \Xe & $> 1.1\times 10^{22}$ & KamLAND-Zen & \cite{Gando2012}       & 12.5 & 0.0011   & $< 1.23$ \\
        \Xe & $> 5.1\times 10^{22}$ & EXO-200     & \cite{Kharusi2021}     & 12.5 & 0.0011   & $< 0.84$ \\
        
        \phibb ($\epsilon_{RR}$) \\ \addlinespace
        \Xe & $> 3.7\times 10^{24}$ & EXO-200 & \cite{Kharusi2021} & --  & -- & -- \\
        \phibb ($\epsilon_{RL}$) \\ \addlinespace
        \Xe & $> 4.1\times 10^{24}$ & EXO-200 & \cite{Kharusi2021} & --  & -- & -- \\
        \bottomrule
    \end{tabular}
    \caption{Comparison of the results obtained by different double-\Beta decay experiments with different isotopes in the search for Majorons-involving decays. The lower limits on the half-life are converted into upper limits on the neutrino-Majoron coupling constant using equation~\ref{eq:majoron_conv} with the axial vector coupling constant $g_A=1.27$ and the phase space factors from~\cite{Kotila2015}. The NME calculations for the spectral index $n=1$ are taken from~\cite{Agostini:2022zub} and references therein, and for $n=3$ and $n=7$ from~\cite{Kotila2021}.}
    \label{tab:comparison_majo}
\end{table*}

A summary of the latest results obtained by different double-\Beta decay experiments is presented in table~\ref{tab:comparison_majo}. 
The most stringent limits, regardless of the isotope and the experiment, are obtained for the model corresponding to a spectral index $n=1$. In fact, the energy distribution predicted for this decay differs the most from the SM \nnbb decay compared to other spectral indexes, as shown in figure~\ref{fig:shape_nnbb_Majorons}. Among different experiments, the best limit on the half-life of the \Jbb decay ($n=1$ mode) is obtained by the EXO-200 experiment with \Xe: $4.3\cdot 10^{24}$\,yr at 90\% \cl. EXO-200 also obtained the best limits on the half-life of the \Jbb/\JJbb decays ($n=3$ modes): $6.3\cdot 10^{23}$\,yr. For the \Jbb decay ($n=2$ mode), the KamLAND-Zen experiment set the most competitive limit on the half-life at $1.0\cdot 10^{24}$\,yr, while the GERDA experiment set the most competitive limit on the half-life of the \JJbb decay ($n=7$ mode): $1.0\cdot 10^{23}$\,yr. 
Recently, EXO-200 has searched for BSM double-\Beta decays in which a Majoron-like particle is emitted via non-standard right-handed currents were also investigated~\cite{Kharusi2021}. Limits were derived on such a decay for both right-handed and left-handed hadronic currents: $3.7\cdot 10^{24}$\,yr and $4.1\cdot 10^{24}$\,yr, respectively.

\paragraph{Lorentz-violating \nnbb decay}

\begin{table*}
    \centering
    \renewcommand{\arraystretch}{1.3}
    \begin{tabular}{cccc}
    \toprule
    Isotope & \alv (GeV) & Experiment & Ref. \\ 
    \midrule
    \Ge & ($-2.7 < \alv < 6.2$) $\cdot 10^{-6}$            & GERDA   & \cite{Agostini2022} \\
    \Se & $\alv < 4.1 \cdot 10^{-6}$                       & CUPID-0 & \cite{Azzolini2019} \\
    \Mo & ($-4.2 < \alv < 3.5$) $\cdot 10^{-7}$            & NEMO-3  & \cite{Arnold2019}   \\
    \Cd & $\alv < 4.0 \cdot 10^{-6}$                       & Aurora  & \cite{Barabash2018} \\
    \Xe & $-2.65 \cdot 10^{-5} < \alv < 7.6 \cdot 10^{-6}$ & EXO-200 & \cite{Albert2016}   \\
    \bottomrule
    \end{tabular}
    \caption{Summary of the results obtained by different double-\Beta decay experiments in the search for Lorentz violation.}
    \label{tab:lorentz_comparison_2nbb_experiments}
\end{table*}

A summary of the latest results obtained by different double-\Beta decay experiments is presented in table~\ref{tab:lorentz_comparison_2nbb_experiments}. 
The most stringent limit on \alv comes from NEMO-3 and is at the order of $10^{-7}$, a factor 10 better than all the other experiments. This result is attributed to the much larger statistics of \nnbb decay events achieved by the NEMO-3 experiment ($\sim 1.9 \cdot 10^5$ events in the analysis range). 
However, part of this difference comes from the statistical treatments as only the CUPID-0 and GERDA experiments used the approach highlighted in this review for class III models (see section~\ref{sec:analysis_methods}), to which the Lorentz violating \nnbb decay belongs. Aurora, NEMO-3, and EXO-200 treated the perturbation introduced by Lorentz violation as an independent component in the fit, neglecting any correlation with the SM \nnbb decay distribution in the result. Nevertheless, the impact is hard to quantify and goes beyond the scope of this work. 
When comparing different results, it should also be considered that the limits on \alv depend on the calculated phase space ratio between the SM \nnbb decay and the LV perturbation. The different experiments reviewed in table~\ref{tab:lorentz_comparison_2nbb_experiments} used different calculations. In recent work, improved phase space calculations were performed, in which the Fermi functions are built with exact electron wave functions obtained by numerically solving a Dirac equation in a realistic Coulomb-type potential, including finite nuclear size and screening effects~\cite{Nitescu:2020xlr}. Differences up to 30\% for heavier nuclei were found between these improved calculations and the previous calculations using approximated analytical Fermi functions. As pointed out in~\cite{Nitescu:2020xlr}, for example, there is a relevant difference between the newly calculated phase space ratio and the one used by the CUPID-0 collaboration in~\cite{Azzolini2019}.

\paragraph{\nnbb decay with bosonic neutrinos}
The experimental search for an admixture of bosonic and fermionic neutrinos through the search for distortions of the \nnbb decay spectrum has been performed only by the NEMO-3 experiment with \Mo~\cite{Arnold2019}. They obtained an upper limit on the bosonic neutrino contribution $sin^2 \chi < 0.27$ at 90\% \cl. 
Also in this case, the statistical treatment  does not follow the receipt given in~\ref{sec:analysis_methods} for class II models as it neglects the correlation between the SM \nnbb decay (fermionic neutrinos) and \nnbb decay with bosonic neutrinos distributions.  
As already introduced in~\ref{subsec:bosonic_nu}, searches with other isotopes than \Mo might be disfavoured by the small predicted ratio $r_0$ with which the fermionic and bosonic contributions are weighted in the total \nnbb decay rate.

\paragraph{Double-\Beta decay into sterile neutrinos and \Ztwo-odd fermions}
The experimental search for light exotic fermions, {\it i.e.} sterile neutrinos and \Ztwo-odd fermions, has been performed only by the GERDA experiment with \Ge~\cite{Agostini2022}, which set a limit on the mixing between active and sterile neutrinos $\sinT <0.013$ for a sterile neutrino mass of $m_N = 500$\,keV. The limits get worse for lower and higher masses ($\sinT < 0.15$ for $m_N=100$\,keV, $\sinT < 0.050$ for $m_N=900$\,keV). GERDA has also set the first direct experimental constraints on the emission of two \Ztwo-odd fermions, constraining the half-life of the corresponding decay to $2.5\cdot 10^{23}$\,yr for a mass $m_\chi=400\,$keV, which translates into a constraint on the coupling $g_\chi$ of $(0.6-1.4)\cdot 10^{-3}$\,MeV$^{-2}$. Again, limits get worse for smaller and larger masses.

\section{Outlook and prospects}\label{sec:prospects_future_searches}

Hunting for the extremely rare \onbb decay, existing experiments collected up to millions of \nnbb decay events. This statistics is expected to rapidly increase, as future experiments, with much larger target mass, will start taking data. We showed in this review that \nnbb decays could be used as powerful probes of new physics. 

Since the first experimental searches for double-\Beta decay with the emission of one or two Majorons, we have seen remarkable progress both in the theoretical description of the decays and in the experimental technologies. Improved and more precise calculations of the phase space factors and NMEs are available today, which are essential to convert the experimental constraints on the half-life of the decays into a coupling between the exotic particle, {\it i.e.} the Majoron, and neutrinos. On the other hand, the experiments reached incredible precision in the study of \nnbb decay with large statistics data samples and drastic reduction of the background compared to their predecessors, pushing the bounds on the half-life of these decays up to $10^{24}$\,yr. 
Limits on the neutrino-Majoron coupling $g_J$ for the Majoron model leading to $n=1$ are also available from astrophysics. Supernova observations allow excluding the region of the parameter space $4\cdot 10^{-7} < g_J < 2\cdot 10^{-5}$ by studying the role of Majorons in the Supernova explosion~\cite{Kachelriess:2000qc, Farzan2002}. Current double-\Beta decay experiments completely excluded the region above the lower Supernova bound. The combined results bring the upper bound down to $g_J < 4\cdot 10^{-7}$, far from the sensitivity of any future double-\Beta decay experiment. Nevertheless, one should remember that Supernova bounds are model-dependent and rely upon additional assumptions. To date, double-\Beta decays provide the best direct constraints on the neutrino-Majoron coupling.

Light exotic fermions can also be searched in double-\Beta decays. Depending on the \qbb of the double-\Beta decay isotope, one or two exotic fermions with a mass between a few hundred keV and a few MeV can be emitted in double-\Beta decay. 
In the search for sterile neutrino in this mass range, current double-\Beta decay experiments provide bounds that are still weaker than the existing single-\Beta decay bounds, as predicted in~\cite{Bolton2020, Agostini2020} and confirmed by GERDA results~\cite{Agostini2022}. Still, future experiments could reach unexplored regions of the parameter space, down to $\sinT \sim 10^{-3} - 10^{-4}$, for masses between 100\,keV and 2000\,keV. To date, no experiment exists or is planned with the capability of testing this part of the parameter space ($100\,\text{keV}< m_N < 2000\,\text{keV}$).\footnote{See figure 7 in~\cite{Bolton:2019pcu}.} Therefore, future double-\Beta decay experiments will provide the best direct constraints on the active-sterile neutrino mixing in the aforementioned mass range. 
In addition, double-\Beta decay experiments offer a unique opportunity to test all those models in which only the double production of light exotic fermions is allowed, leading to the best direct constraints on the coupling between these exotic fermions and neutrinos of the order of $g_\chi \sim 10^{-4}$\,MeV$^{-2}$. 

Studying the \nnbb decay spectrum can provide a sensitive test of the so-called counter-shaded Lorentz violation. Current experiments constrained the Lorentz-violating coefficient \alv at the level of $|\alv| < 10^{-6} - 10^{-7}$\,GeV. The study of the single-\Beta decay spectrum also provides a sensitive constraint to the same coefficient \alv. In~\cite{Diaz2013}, a constraint on \alv was derived using tritium \Beta decay data from the Mainz and Troitsk experiments: $|\alv|< 2.0\cdot 10^{-8}$\,GeV, which is already more competitive of the constraints from double-\Beta decays. Recently, the KATRIN experiment performed a similar analysis using a small set of available data, setting a limit at $|\alv|< 3.0\cdot 10^{-8}$\,GeV~\cite{KATRIN:2022qou}. This limit is expected to further improve up to a sensitivity of $10^{-9}$\,GeV or more with the full KATRIN exposure~\cite{Lehnert:2021tbv}. 
Future double-\Beta decay experiments will be able to improve their current limits (in the best case scenario by a factor of $\sqrt{1/\mathcal{E}}$), nevertheless, hardly reaching single-\Beta decay experiments sensitivity. 

A purely bosonic neutrino would substantially change the total double-\Beta decay rate, therefore, the measured \nnbb decay half-life values. Several precision measurements of the \nnbb decay half-life of different isotopes completely ruled out thy hypothesis of a purely bosonic neutrino. On the other hand, experimental data does not completely exclude the hypothesis of a mixed statistic with a partly bosonic neutrino. The only experimental upper limit on the admixture of the bosonic component was set by the NEMO-3 experiment with \Mo. The sensitivity to spectral distortions depends on the ratio $r_0$ between the rates for purely bosonic and purely fermionic neutrinos, which involves phase space factors and NMEs of the two decays. Consequently, some isotopes are favored ({\it i.e.} \Mo) for future searches of bosonic neutrino admixture compared to others ({\it i.e.} \Ge), for which the very small value of $r_0$ neutralizes possible effects induced by a partly bosonic neutrino. 

Finally, \nnbb decays can be used as a probe of hidden non-standard interaction of neutrinos, like strong neutrino self-interactions and the presence of right-handed currents in weak interactions. Although it was shown that already current experiments could be competitive in constraining such non-standard operators~\cite{Deppisch2020, Deppisch2020b}, no experimental searches have been performed yet.

\begin{acknowledgments}
This work has been supported in part by the German Federal Ministry for Education and Research (BMBF), the German Research Foundation (DFG), the Collaborative Research Center SFB1258,
and by the Science and Technology Facilities Council, part of U.K. Research and Innovation (Grant no. ST/T004169/1). M.A. also acknowledges the support of the UCL Cosmoparticle Initiative.
\end{acknowledgments}

\bibliographystyle{ieeetr}
\bibliography{sample}

\end{document}